 \documentclass[preprint]{aastex}

\usepackage{natbib}
\usepackage{lscape}
\usepackage{longtable}
\usepackage{amsmath,amssymb} 
\usepackage{colortbl} 
\usepackage{subfig}
\usepackage{rotating}
\definecolor{lightgold}{rgb}{1,0.93,0.55}

\shorttitle{IceCube Time-Dependent Searches}
\shortauthors{R.~Abbasi et al.}

\begin{document}


\title{Time-Dependent Searches for Point Sources of Neutrinos with the 40-String and 22-String Configurations of IceCube}


\author{
IceCube Collaboration:
R.~Abbasi\altaffilmark{1},
Y.~Abdou\altaffilmark{2},
T.~Abu-Zayyad\altaffilmark{3},
J.~Adams\altaffilmark{4},
J.~A.~Aguilar\altaffilmark{1},
M.~Ahlers\altaffilmark{5},
K.~Andeen\altaffilmark{1},
J.~Auffenberg\altaffilmark{6},
X.~Bai\altaffilmark{7},
M.~Baker\altaffilmark{1}
% ...
}
\altaffiltext{1}{Department of Physics, University of Wisconsin, Madison, WI 53706, USA}
\altaffiltext{2}{Department of Physics and Astronomy, University of Gent, B-9000 Gent, Belgium}
\altaffiltext{3}{Department of Physics, University of Wisconsin, River Falls, WI 54022, USA}
\altaffiltext{4}{Department of Physics and Astronomy, University of Canterbury, Private Bag 4800, Christchurch, New Zealand}
\altaffiltext{5}{Department of Physics, University of Oxford, 1 Keble Road, Oxford OX1 3NP, UK}
\altaffiltext{6}{Department of Physics, University of Wuppertal, D-42119 Wuppertal, Germany}
\altaffiltext{7}{Bartol Research Institute and Department of Physics and Astronomy, University of Delaware, Newark, DE 19716, USA}
% ...



\begin{abstract}

This paper presents four searches for flaring sources of neutrinos
using the IceCube neutrino telescope. For the first time, a search is performed 
over the entire parameter space of energy, direction and time with sensitivity to neutrino 
flares lasting between 20 $\mu$s and a year duration from astrophysical sources.
Searches which integrate over time are 
less sensitive to flares because they are affected by a larger background of 
atmospheric neutrinos and muons that can be reduced by the use of additional 
timing information. Flaring sources considered here, such as active galactic 
nuclei, soft gamma-ray repeaters and gamma-ray bursts, are promising candidate 
neutrino emitters.

Two searches are "untriggered" in the sense that they look for any possible flare 
in the entire sky and from a predefined catalog of sources from which photon 
flares have been recorded. The other two searches are triggered by multi-wavelength 
information on flares from blazars and from a soft gamma-ray repeater. One 
triggered search uses lightcurves from Fermi-LAT which provides continuous 
monitoring. A second triggered search uses information where the flux states have 
been measured only for short periods of time near the flares. The untriggered 
searches use data taken by 40 strings of IceCube between Apr 5, 2008 and May 20, 
2009. The triggered searches also use data taken by the 22-string configuration 
of IceCube operating between May 31, 2007 and Apr 5, 2008. The results from all 
four searches are compatible with a fluctuation of the background.

\end{abstract}






\section{Introduction} 
\label{sec1}

High-energy neutrinos can be produced in the direct vicinity of charged cosmic ray sources by the interaction of the high-energy cosmic rays with matter or photon fields. In those processes, charged pions are produced and result in a flux of high-energy neutrinos. The latter are unique messengers with which to observe the universe, as they have no charge and interact weakly, traveling directly from their point of creation essentially without absorption, distinguishing them from cosmic rays and high energy photons. Neutrinos are key in understanding the mechanisms of cosmic ray acceleration, and their detection from an astrophysical point source would be a clear indication of hadronic acceleration in that source. Time-integrated analyses suffer from a high irreducible background of atmospheric neutrinos and atmospheric muons, making them less sensitive for the detection of flares. Time-dependent analyses aim to reduce this background by searching over smaller time scales around the flares. The searches discussed in this paper are about a factor of four to five more powerful than time-integrated searches for flares of $\sim 1$~second.

Four searches for time-dependent neutrino emissions from various categories of 
flaring sources are presented in this paper using data from the incomplete 22 and 
40-string configurations of IceCube. We call “triggered” searches those using 
multi-wavelength (MWL) information from photon experiments as a method of 
selecting flaring periods. Based on this information, we select catalogues of
interesting candidate neutrino-emitting flares. We focus on flares of
larger duration than GRBs that are covered by other IceCube searches 
\citet{Abbasi:2009ig}, \citet{Abbasi:2009kq}, \citep{Kevin}). Sources 
we consider are Active Galactic Nuclei (AGN) and one period of activity from the 
newly discovered Soft Gamma Repeater, SGR 0501+4516.

The underlying assumption of triggered searches is that 
the neutrino emission follows the time-dependent emission of the photons, as a 
consequence of an excited state of the source when the jet can accelerate particles 
to higher energies than in the quiescent state. This hypothesis is assumed in the likelihood method as a prior.
In order to make the search for flares as general as possible, an 
``untriggered'' scan for clusters of events in time and direction without prior 
timing information is also performed. Untriggered searches are capable of detecting 
flares of duration similar to Gamma-Ray Bursts (GRBs) to flares lasting many 
days, as seen in AGNs. While dedicated GRB searches are more sensitive, the 
untriggered all-sky search is capable of detecting 'quenched' GRBs which may 
have been undetected by photon telescopes and hence not included in the 
dedicated searches. All-sky searches are affected by large trial factors, hence 
a catalogue of promising sources which are variable in MWL observations is 
selected for an additional untriggered search.

The paper is structured as follows.
In Section \ref{sec2} the properties of the flaring sources, AGNs, Gamma-Ray Bursts (GRBs) and SGRs are considered.
In Section \ref{sec3} the data samples of 40 strings and 22 strings of IceCube are described.
In Section \ref{sec4} the time-dependent likelihood method is illustrated in general and compared with the time-integrated method. 
The four searches for flares are:
\begin{itemize}
\item An untriggered all-sky scan for short-duration neutrino emission from point-like sources (Section \ref{sec6});
\item An untriggered search for flares from a predefined catalogue of 40 sources identified as variable in GeV photons (Section \ref{sec5});
\item A triggered search for continuously monitored sources using MWL information. Data from the Fermi-LAT and also from SWIFT (Section \ref{sec7}) are used;
\item A triggered search using sporadic information on flares collected by various X-ray experiments and Imaging Cherenkov Telescopes (IACTs) with typical energy thresholds around 300 GeV (Section \ref{sec8}). These experiments have a reduced duty cycle compared to Fermi and tend to monitor sources mainly during their flaring states.
\end{itemize}
In sections 5 to 8 each search method and its expected discovery potential are presented, and results are provided. In all searches the background is estimated directly from scrambled data, since the signal contribution is expected to be small. To avoid any bias toward discovery, each search has been performed in a blind fashion by defining cuts and search methods before looking at the final event sample. The final significance, which accounts for the different trials, is calculated by scrambling data in time. For each search the resulting probabilities ($p$-values) that the data could be due to background fluctuation are provided (see also Section \ref{sec4}). 
Trial factors between searches are not included in the $p$-values, as these trial factors are negligible compared to the 5$\sigma$ significance required by IceCube for a claim of discovery.

\section{Candidate Sources of Flaring Neutrino Emission}
\label{sec2}

Galactic and extra-galactic sources exhibit time-dependent variability ranging from short bursts 
with durations between seconds and minutes (e.g. GRBs or giant flares from SGRs) to long periods of high activity lasting hours to weeks (e.g. AGN flares).
One of the main targets of the searches presented in this paper is variable emission from AGNs. Flat-Spectrum Radio Quasars (FSRQs) and BL Lacs, commonly unified in the AGN class of blazars, exhibit relativistic jets pointing towards the Earth and some of the most violent variable high energy phenomena. Their spectral energy distributions (SEDs) extend orders of magnitudes across the electromagnetic spectrum and are characterized by structures of low and high energy non-thermal peaks.
After 11 months of operation, the Fermi-LAT collaboration published their first AGN catalogue \citep{Abdo:2010ge} containing 709 GeV-sources associated with AGNs, many of which are in the previously published Bright Source list catalogue \citep{Abdo:2009mg}. A previous stacking search for neutrinos from AGNs used AMANDA data \citep{Achterberg:2006ik}.

The low energy component in the radio to soft X-rays is due to synchrotron radiation of electrons gyrating in a magnetic field. The high energy component (X-ray to $\gamma$-ray) is explained in leptonic models by synchrotron emissions of electrons in the jet and subsequent up-scattering of photons (inverse Compton) by the electron population responsible for the synchrotron emission. Hadronic models are used to explain the high energy component with relativistic protons, with energies above the threshold for p-$\gamma$ or p-p pion production, which will decay to $\gamma$-rays and also neutrinos. Proton synchrotron emission can also contribute to the high-energy component if they are accelerated to very high energies (for a review on models see e.g. \citet{Ribordy:2009jj,Boettcher:2010wy,Boettcher:2006pd} and references therein).

The emission from blazars is known to be variable at all wavelengths. Simultaneous MWL observations are crucial for understanding the cause of this variability \citep{Gaidos:1996ss, 
Blazejowski:2005ih,Kartaltepe:2007px,Horan:2009vd,Boettcher:2008tq}. The intensity of these objects can vary by more than an order of magnitude between different observing epochs. The typical time scales of AGN flares vary from hours to days, though high-energy variability has been observed on much shorter time scales, in some cases even down to just a few minutes \citep{Aharonian:2007ig,Albert:2007zd}.

While leptonic models enjoy relatively good success reproducing the observed 
emission, other arguments favor a hadronic component. Perhaps the most compelling 
evidence are observations of ``orphan'' flares, defined as TeV photon emission 
without accompanying X-rays, such as the 1ES 1959+650 flare in 2002 
\citep{Krawczynski:2003fq}. An a posteriori observation with AMANDA-II of two 
events \citep{flareamanda}, one exactly during the flare and another 31 days 
later, triggered some theoretical calculations \citep{Halzen:2005pz,Reimer:2005sj}. 
Two recent flares included in the MWL triggered searches (see Sections \ref{sec7} 
and \ref{sec8} also \citep{2008ATel.1650....1C,Ciprini:2009kr,collaboration:2009gga,Vittorini:2008nf}) 
are suspected to be orphan flares, but X-ray observations were not simultaneous 
with gamma-ray observations and there is a possibility of having missed the 
X-ray flare.

GRBs, believed to be produced by the most powerful phenomena in the universe 
\citep{Meszaros:2006rc,Piran:2004ba}, are interesting as time-dependent candidate 
neutrino sources \citep{Waxman:2003vh,Meszaros:1993}. IceCube conducts dedicated 
searches which are triggered by satellite information for these objects 
(\citet{Abbasi:2009ig}, \citet{Abbasi:2009kq}, \citep{Kevin}). The untriggered 
all-sky search presented in this paper is also sensitive to this source class if 
two or more neutrinos can be detected from the same GRB.  While the dedicated 
searches are in general much more sensitive (because they use the known time and 
direction of GRBs observed in gamma- or X-rays), the untriggered search has the 
potential to detect a burst which was not observed in photons (due to e.g. 
absorption or lack of monitoring).

Another possibility for powerful emission is given by SGR's, X-ray pulsars that 
show variability at different timescales and a persistent X-ray emission with 
luminosity $L \sim 10^{35}$~erg/s with short bursts of X- and $\gamma$-rays with 
$L \sim 10^{41}$~erg/s  lasting $\sim 0.1-1$ seconds (for review see \citet{Mereghetti:2008je}).  
These X-ray pulsars, together with Anomalous X-ray Pulsars, are considered to 
be the best candidates for magnetars, isolated neutron stars powered by huge 
magnetic fields ($B\sim10^{15}~G$).  
At times these sources emit giant flares with initial spikes of hard non-thermal
radiation up to luminosities of $\sim 10^{46}$~erg/s lasting some seconds. The 
enormous photon fluxes from these flares saturate most detectors. The flares may 
also accelerate baryons and produce neutrinos \citep{Halzen:2005tc,Ioka:2005er,Liu:2009aj}. 
Limits for photons in the 10 TeV-100 PeV energy range using AMANDA-II data were 
published from the powerful giant flare observed in Dec. 2004 from SGR 1806-20 
\citep{Achterberg:2006az}.
In the catalogue (Table \ref{ic40flare}) used in one 
of the triggered flare searches (Section \ref{sec8}), we consider six days from 
a period of intense flaring 
from SGR 0501+4516 is included, which was discovered by SWIFT beginning Aug. 22, 2008 and also observed 
by RXTE/ASM, Konus-Wind and the Fermi GBM \citep{Kumar:2010fh}. 
Even if the period of activity of SGR 0501+4516 is not as intense as a giant flare 
and the observed power law component was quite soft \citep{Rea:2009nh}, this was 
a long-lasting period of flares that lead to the object's discovery as an SGR. 
Periods of activity from known SGRs 1E 1514-57 and SGR 1806-20 were omitted, since 
the large background of atmospheric muons severely limits IceCube sensitivity to
southern hemisphere objects.


\section{The IceCube Detector and the Data Sample for this Analysis}
\label{sec3}

\begin{figure}[ht]
 \centering{
\begin{tabular}{c}
    \includegraphics[width=3.5in,height=3.3in]{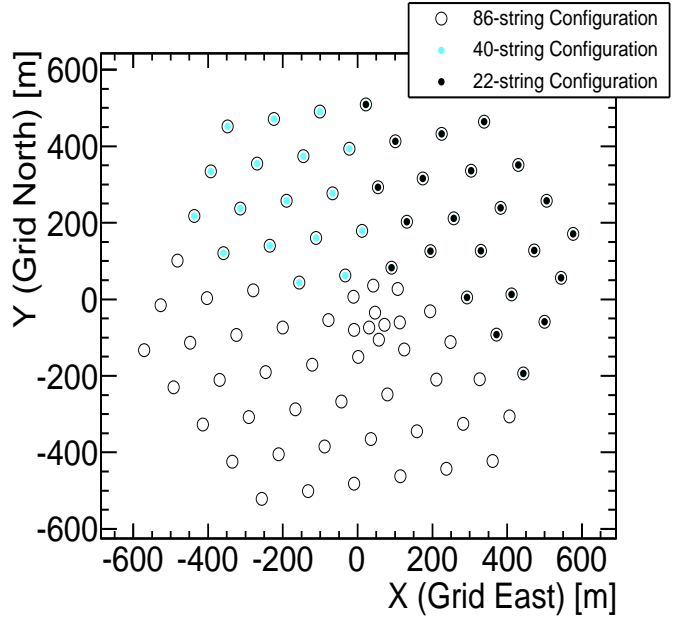}
 \end{tabular}
 }
 \caption{The growing IceCube detector seen from the top. Filled circles inside empty circles indicate deployed strings for each configuration, where all strings used in the 22-string configuration were also used in the 40-string configuration.}
 \label{ic2240_geometry}
\end{figure}

The IceCube Neutrino Observatory is composed of a deep array of 86 strings holding 5,160 digital optical modules (DOMs), which are deployed between 1.45 and 2.45 km below the glacial surface at the geographic South Pole \citep{2002RPPh...65.1025H,2008PhR...458..173B}.  
IceCube strings are horizontally separated by about 125 m with DOMs positioned vertically 17 m apart along each string. Each DOM consists of a Hamamatsu photomultiplier with 25 cm diameter \citep{Abbasi:2010vc}, electronics for waveform digitization \citep{IceCubeDAQ_:2008ym}, and a spherical, pressure-resistant glass housing.
IceCube construction started with a first string installed in the 2005-6 season \citep{IceCubeFirstYearPerformance_Achterberg:2006md} and has recently been completed in the austral summer of 2010-11.
The configurations of IceCube that have been used for the present analyses (22-string and 40-string) are shown in Figure~\ref{ic2240_geometry}.
The IceCube Neutrino Observatory includes a dense subarray, DeepCore, designed to enhance the physics performance of IceCube below 1 TeV \citep{Cowen:2009zz} and a surface array, IceTop, for extensive air shower measurements on the composition and spectrum of cosmic rays \citep{collaboration:2010zw}.

IceCube uses a simple multiplicity trigger, which requires that at least eight DOMs are triggered within 5 $\mu$s. For a DOM to trigger, it is both required that the DOM PMT voltage crosses the discriminator threshold (0.25 of a typical photoelectron), and this ``hit'' to be in coincidence with at least one other hit in the nearest or next-to-nearest neighboring DOMs on a string within $\pm 1\mu$s. This greatly reduces hits due to uncorrelated PMT noise and radioactivity in the glass.
Once the simple multiplicity condition is satisfied, information from all triggered DOMs within a $\pm$10 $\mu$s window is read out and merged to create an event. This means that 20 $\mu$s is the effective limit on how close two events can be in time for the 40-string or 22-string data. Improvements in event definition remove this constraint for data taken with the completed detector.
Standard IceCube runs are eight hours long, runs in the 40 and 22-string configurations have a roughly two minute gap between the end of one run and the beginning of the next, improvements have eliminated this gap in the operation of the full detector.
The rate of each run is monitored and checked for any deviation from an average that accounts for a seasonally adjusted average rate \citep{Tilav:2010hj}. 

Identification of neutrino-induced muon events in IceCube was demonstrated using atmospheric neutrinos as a calibration tool \citep{IceCube9StringAtmNu_Achterberg:2007bi}. The measurement of atmospheric neutrinos by the 40-string configuration has been presented in \citep{PhysRevD.83.012001}.
This sample of upgoing events dominated by atmospheric neutrinos is used to look for astrophysical signals from point sources using directional and energy information.  
In the 40-string sample, the IceCube field of view has been extended compared to the upgoing only sample of 22-strings \citep{Abbasi:2009iv} to also include downgoing events from the southern hemisphere. 
This technique was used for the first time to include events to $-50^{\circ}$ declination with the 22-string configuration \citep{Abbasi:2009cv}.
The background sample in the downgoing region consists of very high energy muons. Atmospheric muons are roughly five orders of magnitude more numerous than neutrino-induced muons at the depth of IceCube. However, their number can be reduced by selecting high energy events so that the astrophysical signal can potentially emerge, if the signal spectrum is harder than that of the atmospheric background.
This results in a different sensitivity for the northern hemisphere, where TeV-PeV neutrino astronomy is possible due to absorption of atmospheric muons in the Earth, with respect to the southern hemisphere, where only PeV-EeV neutrino searches are performed in the current analysis.

The selection of the data from the IceCube 40-string configuration used in this search is discussed in the paper on the time-integrated searches, \citet{Jon}. We refer to this paper for all details on the 
selection of the muon events in the filtered stream (L1) that is sent over satellite from the South Pole and on the final cuts to obtain the data sample used in this analysis. 
The cuts were optimized using an $E^{-2}$ spectrum signal, as expected of neutrinos
directly accelerated in astrophysical sources via stochastic shock acceleration. While this idea
was first presented by \citep{PhysRev.75.1169,fermi54}, and further developed by
e.g. \citet{krymskii77,bell78a,bell78b}. The resulting particle spectra follow a
power-law close to $E^{-2}$. Detailed calculations show, however, that
depending on the shock conditions, the spectra can also be somewhat
flatter or steeper, see e.g. \citet{Steckeretal07,melietal08}. Here, we use an $E^{-2}$
spectrum as a first order estimate.

It also includes a detailed discussion on the analysis method. 
The final sample consists of 36,900 atmospheric neutrino and muon events from the whole sky ($-85^{\circ}$ to +85$^{\circ}$ in declination) detected by IceCube in the 40-string configuration in 375.5 days of good data taking, which corresponds to 92\% uptime during the nominal operation period between April 5, 2008 and May 20, 2009 or Modified Julian Date (MJD) 54561--54971. In this sample, 14,121 events are up-going, while 22,779 events are down-going. Deadtime for this analysis is mainly due to test and calibration runs during and after the construction season.
For time-dependent analyses in general, parts of the detector may be excluded for short periods from the acquisition, but the remaining part can be useful in case an astrophysical event occurs (see e.g. \citet{Abbasi:2009kq}).

In the triggered search for flares (Section \ref{sec8}) we also consider events 
observed during data taking with 22 strings of IceCube, with a livetime of 275.7 days, or 89\% of the operation period from May 31, 2007 to April 5, 2008 (MJD 54251-54561). This sample is described in \citet{Abbasi:2009iv}, and consists of 5114 candidate events from declinations -5$^{\circ}$ to +85$^{\circ}$.

The systematic uncertainties have been evaluated and presented in Sec. 6 of \cite{Jon} for the 40-string data and have also been discussed in \citet{Abbasi:2009iv} for the 22-string data. The main uncertainties on the limits to the fluence of an $E^{-2}$ signal of muon neutrinos come from photon propagation, absolute DOM efficiency, and uncertainties in the Earth density profile and muon energy loss, accounting for a total of 16\%. In this paper we focus on issues related to transient searches, such as the stability in time of the data sample and effects of the detector asymmetry for flares lasting less than one day.

\begin{figure*}[ht]
 \centering{
   \includegraphics[width=5.5in]{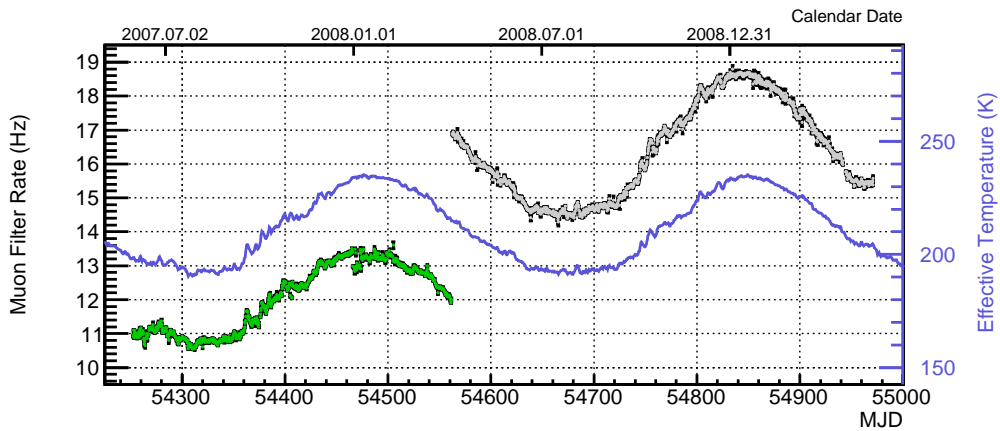}
 }
 \caption{ The rate per run of the filtered stream of muon events with zenith angle $\ge80^{\circ}$ selected at the South Pole for the 22 (dark green points before MJD 54560) and 40 (gray points after MJD 54560) string detectors as a function of MJD. The small modulations around the main seasonal oscillation are due to short-term weather variability (plotted in light blue). }
 \label{fig3}
\end{figure*}

\begin{figure}[!ht]
 \centering{
   \includegraphics[width=4.5in]{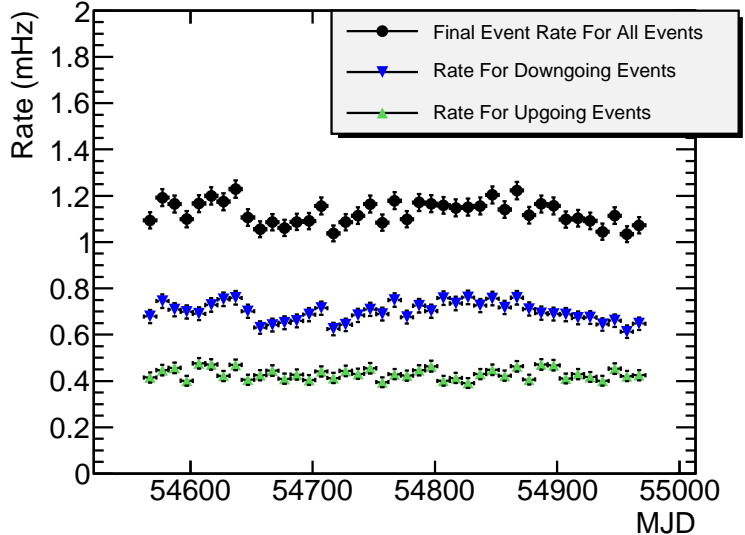}
 }
 \caption{A graph of the rate of the final sample of 40 string events, in bins of 10 days. Errors are statistical. Also plotted are the individual rates of upgoing and downgoing events. The total fluctuation in the final data rate is $\pm 5\%$ for downgoing events and $\pm \sim4\%$ for upgoing events. The analysis in Section \ref{sec5} uses a sinusoidal fit to the atmospheric muon event rate to estimate rate fluctuations in the downgoing region. All other searches neglect these event rate fluctuations, which are negligible compared to statistical fluctuations in the expected signal.}
 \label{figrates}
\end{figure}

Before cuts are applied to the data, the samples are dominated by downgoing atmospheric muons.
This is the case in the upgoing signal region as well, since some atmospheric muons are misreconstructed as upgoing and must be rejected in the process of applying analysis cuts.
The atmospheric muon rate exhibits a seasonal
variation of roughly $\pm 10\%$ due to changes in density of the atmosphere at the South Pole \citep{Tilav:2010hj}. 
When the atmosphere is warmer and less dense during the austral summer, the fraction of pions and kaons in air showers which decay before interacting is increased compared to the winter.
The muon rate also varies several percent on timescales of several days as a result of weather phenomena in Antarctica. 
For upgoing atmospheric neutrinos 
the seasonal variations are smaller, approximately 5\%, since neutrinos are created over a wide range of Earth's latitudes compared to the atmospheric muons created near the South Pole.
To ensure stable detector conditions, the event rates of runs were required to be within $5\sigma$ from a rolling average over $\pm 2$ days.
This loose constraint allows for short-term weather variability. 
All events have initial reconstructions performed using track and cascade based hypotheses, some of which are selected for transmission over satellite to the northern hemisphere for additional processing and analysis.
The muon filter focuses on the selection of upward-going track-like events.
The temporal variation of the event rate for the 22-string and 40-string runs is shown in Figure~\ref{fig3}, where the seasonal modulation is clearly visible.
The rate of the 40 string final event sample is shown in Figure~\ref{figrates} for upgoing and downgoing events and for the total sample.

Due to the requirements for triggering and filtering, the cuts applied, Earth absorption properties and detector geometry, the final sample of events is not uniform in the detector local coordinates zenith ($\theta$), and azimuth ($\phi$).
For time-integrated point source searches, the azimuth dependence is usually neglected because it is smoothed in right ascension by the rotation of the Earth over long integration times.
However, in a time-dependent analysis the azimuth dependence becomes important for time scales shorter than 1 day. 
The local coordinate (zenith and azimuth) distribution of 40 string data is shown in Figure~\ref{figlocalcoords} (left).
In the northern sky there is the effect that events traveling along the longer axis (see Figure~\ref{ic2240_geometry}) of the detector have a longer lever arm, and are more likely to trigger the detector and be well-reconstructed. Well reconstructed events typically have angular reconstruction errors of $<1^{\circ}$ and energy reconstruction resolution of 0.3 in $\log(\mathrm{Energy})$. In the southern sky, there is an online cut on the integrated charge seen in all DOMs for a given event. This gives a preference to events which pass near a line of strings, yielding a six-fold peak in rates corresponding to the main axes of the detector symmetry.

\section{Unbinned Time-Dependent Likelihood Method}
\label{sec4}

The unbinned likelihood searches performed here are based on the method described in \citet{Braun:2008bg_Methods} and extended to searches for time-dependent behavior in \citet{Braun:2009wp}.
In this likelihood ratio method, a combination of signal and background populations is used to model the data.
For a data set with $N$ total events, the probability density of the $i^{th}$ event is given by:

\begin{equation}
\frac{n_s}{N}\mathcal{S}_i + ( 1-\frac{n_s}{N} )\mathcal{B}_i.
\label{llh_event}
\end{equation}

where $n_s$ is the unknown number of signal events with signal fraction of $n_s/N$, $\mathcal{B}_i$ is the background probability density function (PDF) and $\mathcal{S}_i$ is the signal PDF. The likelihood $\mathcal{L}$ of the data given the value of $n_s$ is defined as the product of the individual event probabilities:

\begin{equation}
\mathcal{L}(n_s) = \prod_{i=1}^{N} \Big[\frac{n_s}{N}\mathcal{S}_i + (1 - \frac{n_s}{N})\mathcal{B}_i\Big].
\label{eq:lh}
\end{equation}
This likelihood is maximized with respect to $n_s$ and any other fit parameters which are a part of the signal hypothesis. The maximization provides the best-fit values of these parameters.
The background PDF, $\mathcal{B}_i$, is given by:

\begin{equation}
\mathcal{B}_i = B^{\mathrm{space}}_{i}(\theta_{i},\phi_{i}) B^{\mathrm{energy}}_{i}(E_{i},\theta_{i}) B^{\mathrm{time}}_{i}(t_{i},\theta_{i}),
\label{llh_bg_time1}
\end{equation}
and is computed using the distribution of data itself.

\begin{figure*}[ht]
 \centering{
  \begin{tabular}{cc}
  \includegraphics[width=3.1in]{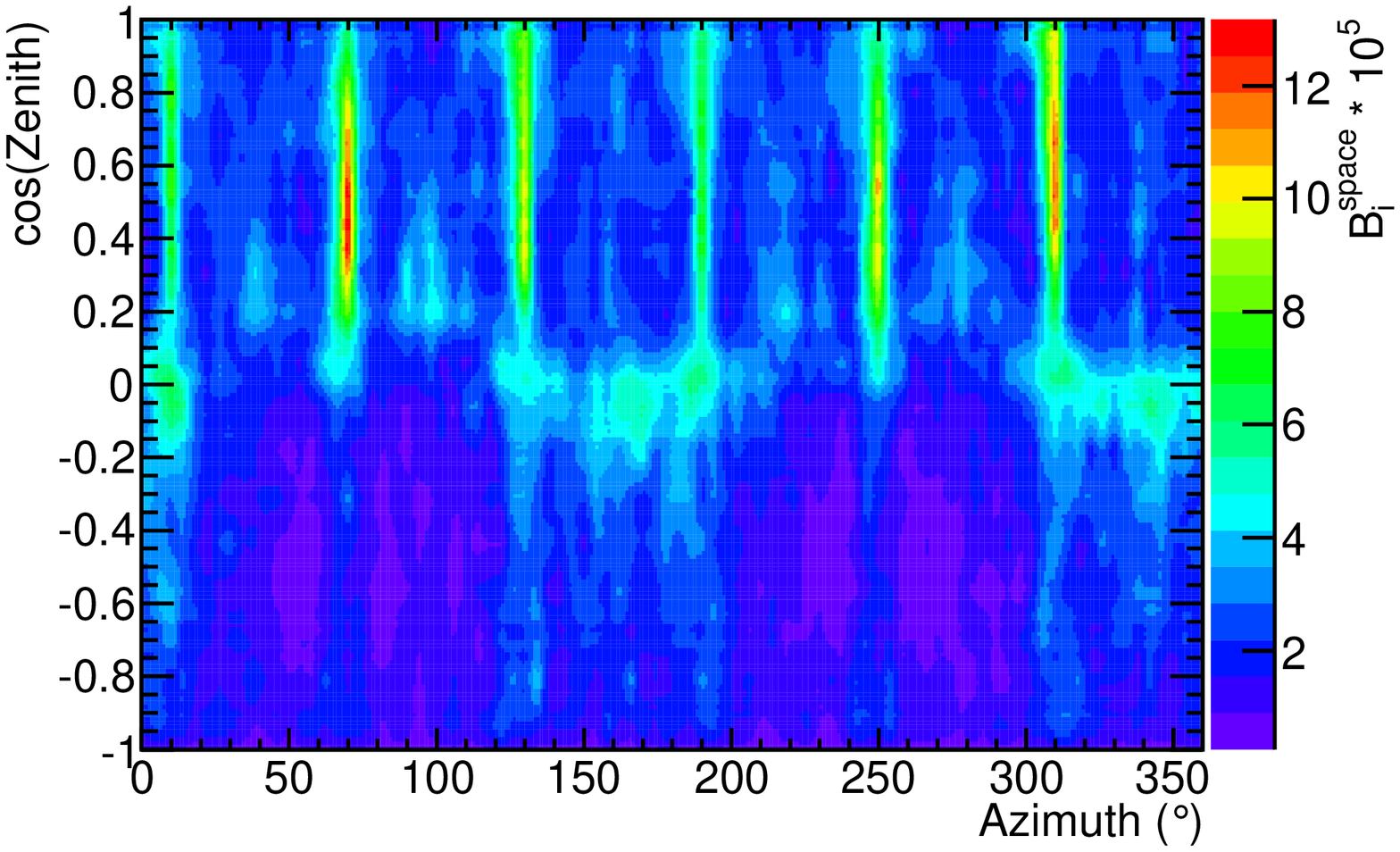} &
  \includegraphics[width=3.1in]{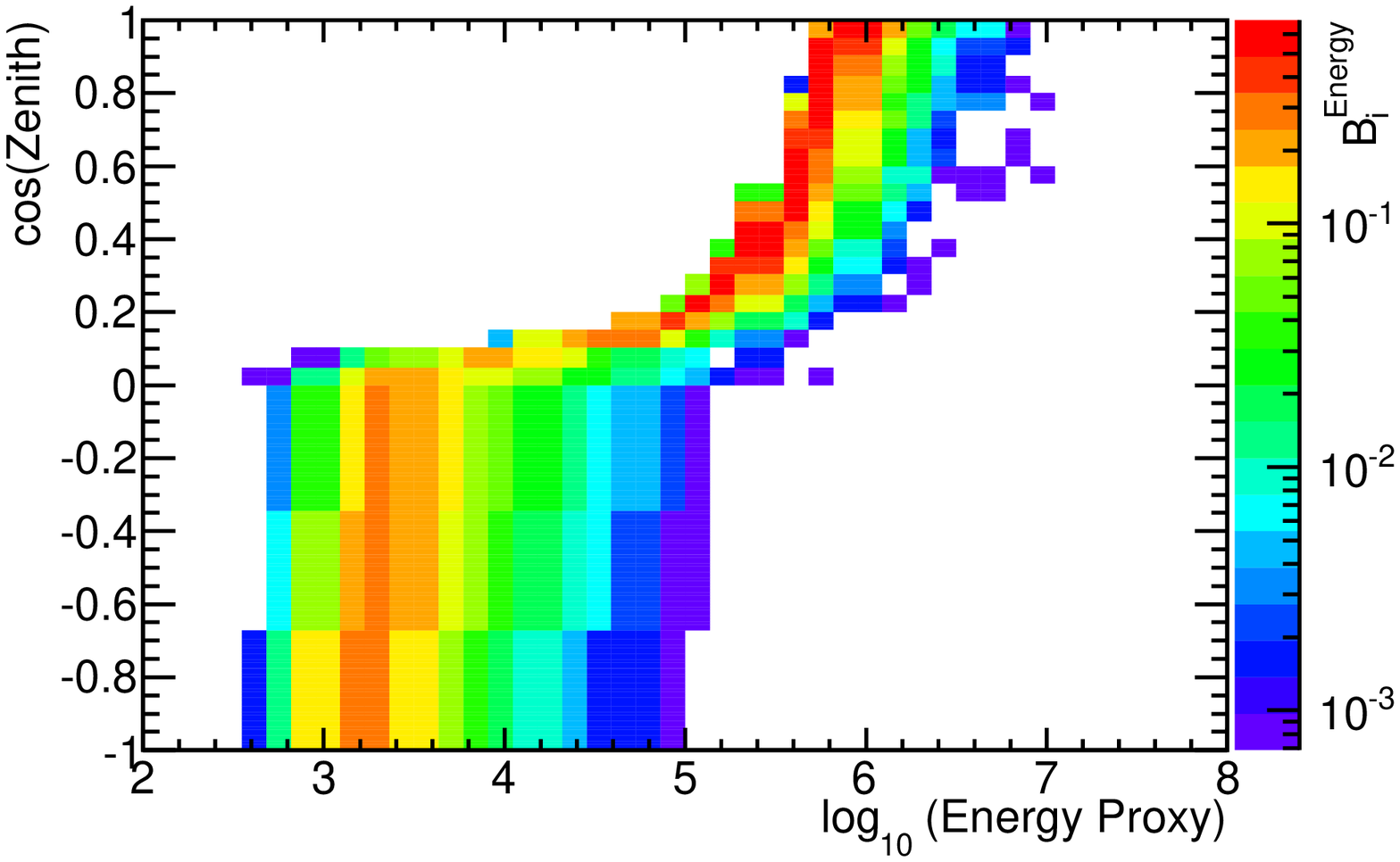}
  \end{tabular}

 }
\caption{Left: The normalized event distribution of the 36,900 events in local coordinates for the 40 strings data (the space term in Equation \ref{llh_bg_time1}). There are two predominant effects: for upgoing events (northern sky, bottom half), events traveling down the longer end of the detector are more likely to trigger and pass cuts; for downgoing events (southern sky, top half), there are six peaks in the event rate. This is due to the initial filter conditions at the South Pole that select tracks more efficiently when they pass close to aligned strings. Right: The background energy distribution (the energy term in Equation \ref{llh_bg_time1}), which is the normalized distribution of event energy values split into bins of the cosine of zenith.
}
\label{figlocalcoords}
\end{figure*}

The spatial term $B^{\mathrm{space}}_{i}(\theta_{i},\phi_{i})$ is the event density per unit solid angle as a function of the local coordinates, shown in Figure~\ref{figlocalcoords} (left). The energy probability, $B^{\mathrm{energy}}_i(E_{i},\theta_{i})$, is determined from the reconstructed energy distribution of data as a function of the cosine of the zenith angle $\cos\theta_{i}$ (see Figure~\ref{figlocalcoords} on the right). 
This energy reconstruction, described in detail in \citet{Jon}, compares the measured to the expected density of photons along the muon track due to stochastic energy losses of pair production, bremsstrahlung and photonuclear interactions which dominate over ionization losses for muons above 1 TeV. The reconstructed energy value represents an estimate of the muon energy in GeV.
The energy cut for the southern sky sample increases for smaller zenith angles, creating a strong zenith dependence of the energy in the southern sky as can be seen in Figure~\ref{figlocalcoords} (right). The goal of the energy cut is to sample a constant number of events per unit solid angle in the southern sky. 
Note that for the northern sky the energy dependence on zenith is small.
The time probability $B^{\mathrm{time}}_{i}(t_{i},\theta_{i})$ of the background can be 
modeled including the expected seasonal modulations (done in Section \ref{sec5}), which are less than $\pm 10\%$ and depend on the zenith angle, or taken to be flat since these modulations are negligible compared to possible signal fluctuations (all other searches).

The signal PDF $\mathcal{S}_i$ is given by:

\begin{equation}
S_{i}=S^{\mathrm{space}}_i(\mid \vec{x}_i-\vec{x}_{s} \mid, \sigma_{i})S^{\mathrm{energy}}_i(E_{i},\theta_{i},\gamma_{s})S^{\mathrm{time}}_{i} ~~,
\label{llh_signal_time1}
\end{equation}
where $S^{\mathrm{space}}_i$ depends on the angular uncertainty of the event 
$\sigma_{i}$ and the angular difference between the event coordinate $\vec{x}_i$ 
and the source coordinate $\vec{x}_{s}$. $S^{\mathrm{energy}}_i$, the energy 
dependent PDF which is a function of the reconstructed event energy $E_{i}$, and 
of the spectral index $\gamma_{s}$ (a power-law spectrum with no cutoff such that 
$dN/dE \propto E^{-\gamma_{s}}$) is calculated from an energy distribution of 
simulated signal in a zenith band that contains the event. $S^{\mathrm{time}}_{i}$ 
is the time-dependent signal PDF. It depends on the particular signal hypothesis, 
which will be described in detail in each section. 
For each search, signal is injected with the same functional form in 
time as is being tested. 
Due to the low number of signal events involved, cross tests using one emission 
profile for the injection and a different one for the search yield largely
similar results. The searches were designed to improve discovery 
potential with a minimum of source assumptions.


The test statistic ($TS$) is calculated from the likelihood ratio of the background-only (null) hypothesis over the signal-plus-background hypothesis: 

\begin{equation}
TS = -2 \log\Big[\frac{\mathcal{L}(n_s=0)}{\mathcal{L}(\hat{n}_s,\hat{\gamma}_s,\hat{T}_s)}\Big].
\label{eq:ts}
\end{equation}
As expressed in Equation \ref{eq:ts}, the test statistic values for scrambled 
samples will distribute as a chi-square function with number
of degrees of freedom equal to the number of fit parameters. The best fit parameters 
$\hat{n}_s,\hat{\gamma}_s$ and any fit time parameters $\hat{T}_s$ (which will be 
explained for each search) are obtained by maximizing $TS$. Several searches differ in the 
methods used to find the maximum value of $TS$, and may include weighting terms 
for values of a best fit parameter, which will be included in their description below.

Larger values of $TS$ are less compatible with the null hypothesis, and indicate its rejection at a significance level corresponding to the fraction of the scrambled trials above the $TS$ value found in the data. 
Data scrambling is done by assigning a random time to each event from a period of active data taking
and performing the proper coordinate transformation to get a new right ascension and declination. 
The fraction of trials above the $TS$ value obtained from data is referred to as the {\em $p$-value}.
This leads to the definition of the discovery potential at a particular significance level: the average number of signal events required to achieve a $p$-value less than a specified threshold in 50\% of trials. IceCube uses a one-sided 5$\sigma$ significance level as the threshold for discovery, corresponding to a $p$-value of 2.87$\times 10^{-7}$. Similarly, the sensitivity is defined as the average signal required to obtain a $p$-value less than that of the median of the test statistic distribution of scrambled (background-only) samples in 90\% of trials.

\begin{figure}[h]
 \centering{
   \includegraphics[width=4.5in]{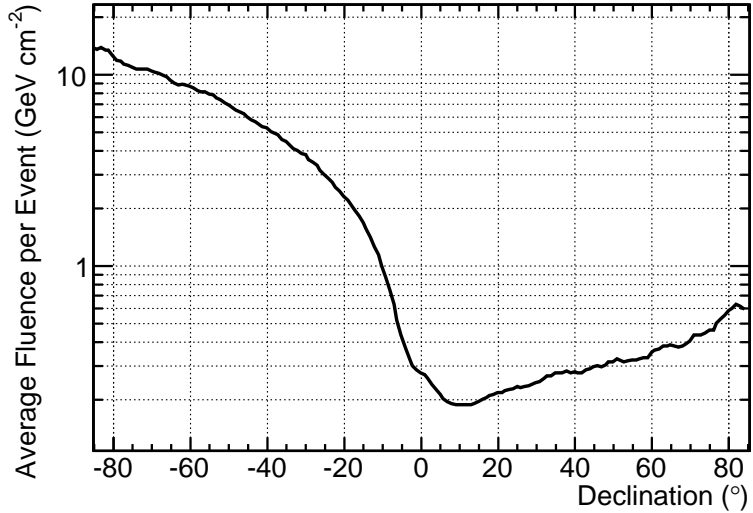}
 }
 \caption{The total fluence from an $E^{-2}$ spectrum muon neutrino signal in a declination band divided by the number of events in the band in the 40-string configuration, plotted against declination. }
 \label{fig:em2fluence}
\end{figure}

Aside from the $p$-values from searches, in the absence of a signal upper limits on the fluence are provided, defined as the integral in energy and time of the flux upper limit:
\begin{equation}
f=\int^{t_{\mathrm{max}}}_{t_{\mathrm{min}}}dt \int^{E_{\mathrm{max}}}_{E_{\mathrm{min}}} dE \times E\frac{dN}{dE}=\Delta t \int^{E_{\mathrm{max}}}_{E_{\mathrm{min}}}dE \times E \frac{\Phi_0}{E^{2}} = \Delta t \Phi_0 \left [ \log(E_{\mathrm{max}})-\log(E_{\mathrm{min}}) \right] ~~,
\label{eq:fluence}
\end{equation}
where $\Phi_0$ is the upper limit on the normalization on an $E^{-2}$ spectrum and $E_{\mathrm{min}}$ and $E_{\mathrm{max}}$ specify the declination dependent energy range containing 90\% of the signal spectrum at final sample selection criteria obtained from simulations. $\Delta t$ is the duration of the emission.
There is a correspondence between the fluence and the average number of events detected, shown as a function of the declination in Figure~\ref{fig:em2fluence}. 
The limits are calculated according to the classical (frequentist) construction of upper limits \citep{Neyman:1937}
and the systematic error of 16\% is neglected in all upper limits since the limits are dominated by statistical fluctuations for flares.

Limits in this paper have been produced assuming a flux of only muon neutrinos. The scenario of standard neutrino oscillations over
astronomical distances \citep{Athar} assumes equal fluxes of all flavors of neutrinos at the Earth from a source producing neutrinos via pion decay with a ratio of $\nu_{e} : \nu_{\mu} : \nu_{\tau} = 1:2:0$.
When considering equal fluxes of muon and tau neutrinos at the Earth, the resulting upper limits on the sum of both fluxes are about a
factor of 1.7 times higher than if only muon neutrinos are considered rather than the expected factor of two due to oscillation if no tau neutrinos would be detectable.
For an $E^{-2}$ spectrum of the signal neutrino flux the contribution due to the detectable tau neutrino flux for sources at the horizon is 10\% and up to 15\% for sources in the Northern hemisphere. This is due to the tau decay channel into muons with a branching ratio of 17.7\% and in part to the tau leptons with energy greater than some PeVs that may travel far enough to be reconstructed as tracks in IceCube before decaying. In the upgoing region we have considered tau regeneration in the Earth.

\section{All-Sky Time Scan}
\label{sec6}

The all-sky time-dependent search presented here complements the all-sky search applied to the IceCube 40-string data in \citep{Jon}. While that search has the best sensitivity to steady sources, a source which has emitted neutrinos for only a limited period of time might not be detected. The time-dependent analysis here scans for a significant excess with respect to background over all time scales (from sub-seconds to the full year) at each direction of the sky. For flares shorter than $\sim$100 days, the discovery potential of the time-dependent search typically becomes better than the time-integrated one, and in principle a short burst can be discovered with only two events if they occur close enough together in time ($\sim0.1$ seconds for $E^{-2}$ spectrum events). The advantage of such untriggered searches is their ability to cover all emission scenarios, including neutrino emission without any observed counterpart in the electromagnetic spectrum.

\subsection{Method and Expected Performance} 

\begin{figure}[h]
 \centering{
   \includegraphics[width=3.5in]{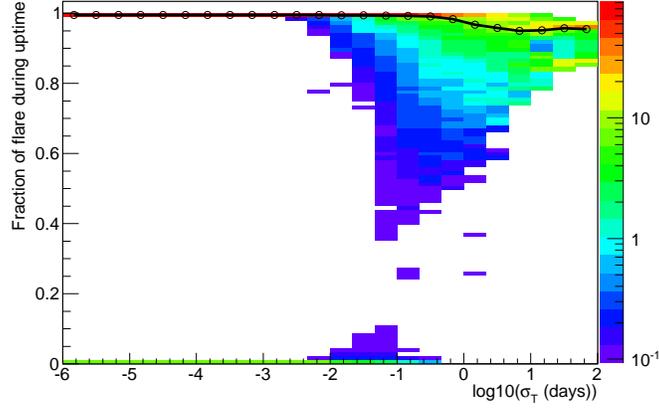}
 }
 \caption{The fractional duration of randomly-simulated flares which occur during the uptime for the 40-string configuration for a range of different flare durations. The black line marks the median fraction of fluence occurring during the detector livetime for a given flare duration, which is used as a correction factor for the fluence of observed flares. For instance, for flares shorter than one minute, there is approximately an 8\% chance of the flare occurring completely during detector downtime. Flares longer than one day will always have some emission during uptime; on average 92\% of the total emission will coincide with usable run time.}
 \label{fig6.0}
\end{figure}

The method in \citet{Braun:2009wp} is adapted for this search to a real detector with non-uniform acceptance and deadtime. The non-uniform acceptance can be seen in \ref{figlocalcoords}, deadtime compensation is shown below in \ref{fig6.0}. The time-dependent probability density function from Equation \ref{llh_signal_time1} for this search is a Gaussian function:

\begin{equation}
S^{\mathrm{time}}_i = \frac{1}{\sqrt{2\pi}\sigma_T} \exp \left(-\frac{(t_i-\text{T}_{\circ})^2}{2 \sigma_T^2}\right)
\end{equation}
where $t_i$ is the arrival time of the event, and fit parameters $\text{T}_{\circ}$ and $\sigma_T$ 
are the mean and sigma of the Gaussian describing flaring behavior in time.
The maximization of the test statistic returns the best-fit values of the 
Gaussian mean (the time at which the flare peaks) and sigma (corresponding to 
the duration of the flare). Both the background and expected number of events 
are small, distinguishing a box-type function from a Gaussian would require many 
more events than required for a 5$\sigma$ discovery, and we find that using 
either of these flare hypotheses performs similarly (see Section~\ref{sec5}). It was found that the fitting
method used in this section worked better with a continuous function, so a Gaussian
functional form was chosen.


Because there are many more independent small time windows than large ones, the 
test statistic formula of Equation \ref{eq:ts} is modified to include a weighting 
term to correct for this effective trial factor and avoid undue preference for 
short flares using a Bayesian approach \citep{Braun:2009wp}. The test statistic 
formula that is maximized is then:

\begin{equation}
  TS = -2 \log\Big[\frac{\text{T}}{\sqrt{2 \pi}\hat{\sigma_T}} \times \frac{\mathcal{L}(n_s=0)}{\mathcal{L}(\hat{n}_s, \hat{\gamma}, \hat{\sigma}_{T}, \hat{\text{T}}_{\circ})}\Big]~,
  \label{marginalize}
\end{equation}
where the first factor in the square brackets is the weighting term and the second is the likelihood ratio. T is the total livetime of data taking, $\hat{n}_s, \hat{\gamma}, \hat{\sigma}_{T}, \hat{\text{T}}_{\circ}$ are the best-fit values for the number of signal events, spectral index, width and mean of the Gaussian flare, respectively. In order to prevent the weighting term from becoming less than 1,
a constraint is placed on the flare width $\sigma_{T}$.
This is done to prevent flares with zero amplitude ($\hat{n}_s$=0) from having a positive test statistic, which would happen if the flare width $\sigma_{T}$ were allowed to be greater than ${\text{T}}/{\sqrt{2 \pi}}$.

\begin{figure}[!ht]
 \centering{
\begin{tabular}{cc}
   \includegraphics[width=3.3in]{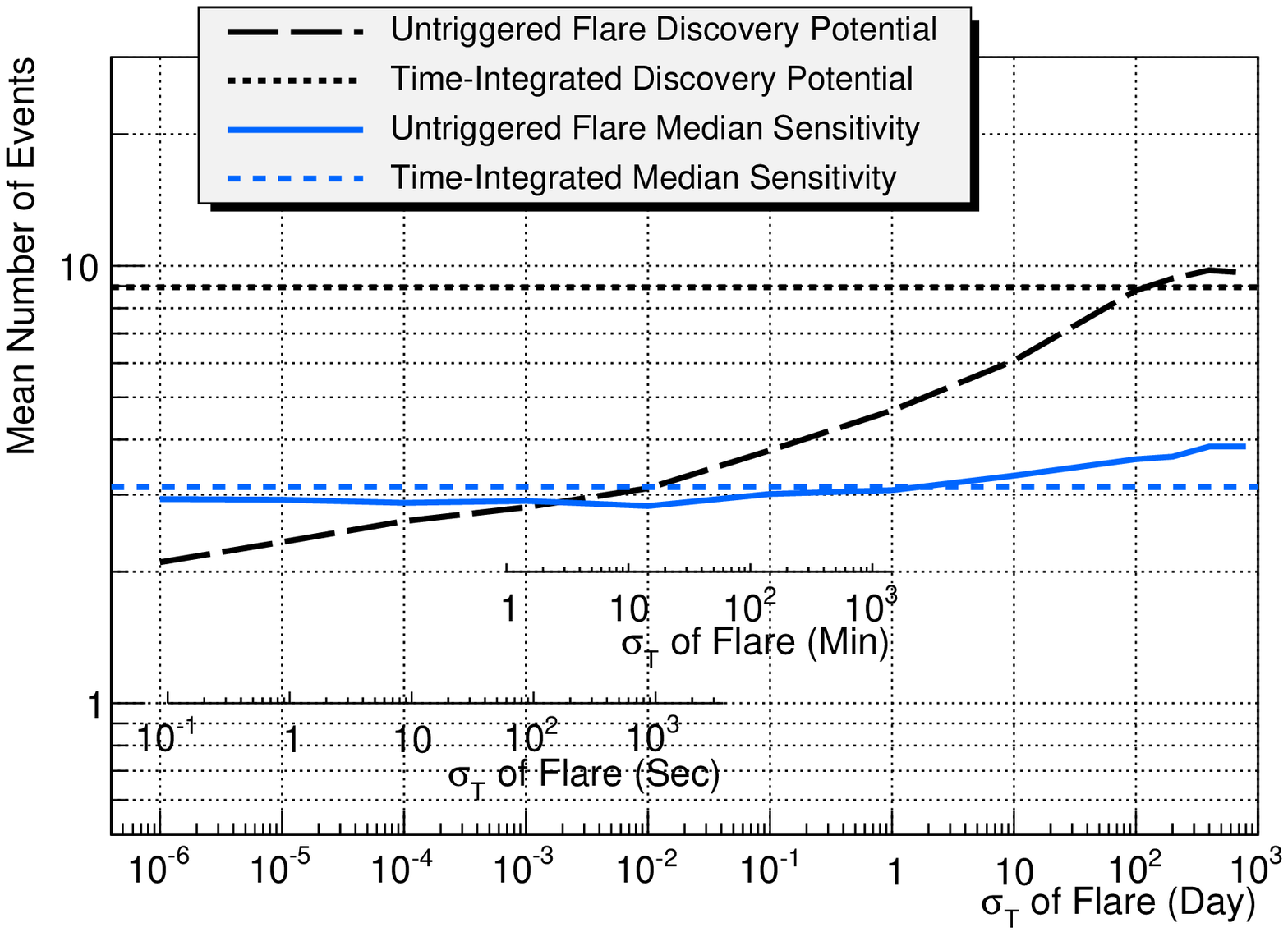} &
   \includegraphics[width=3.3in]{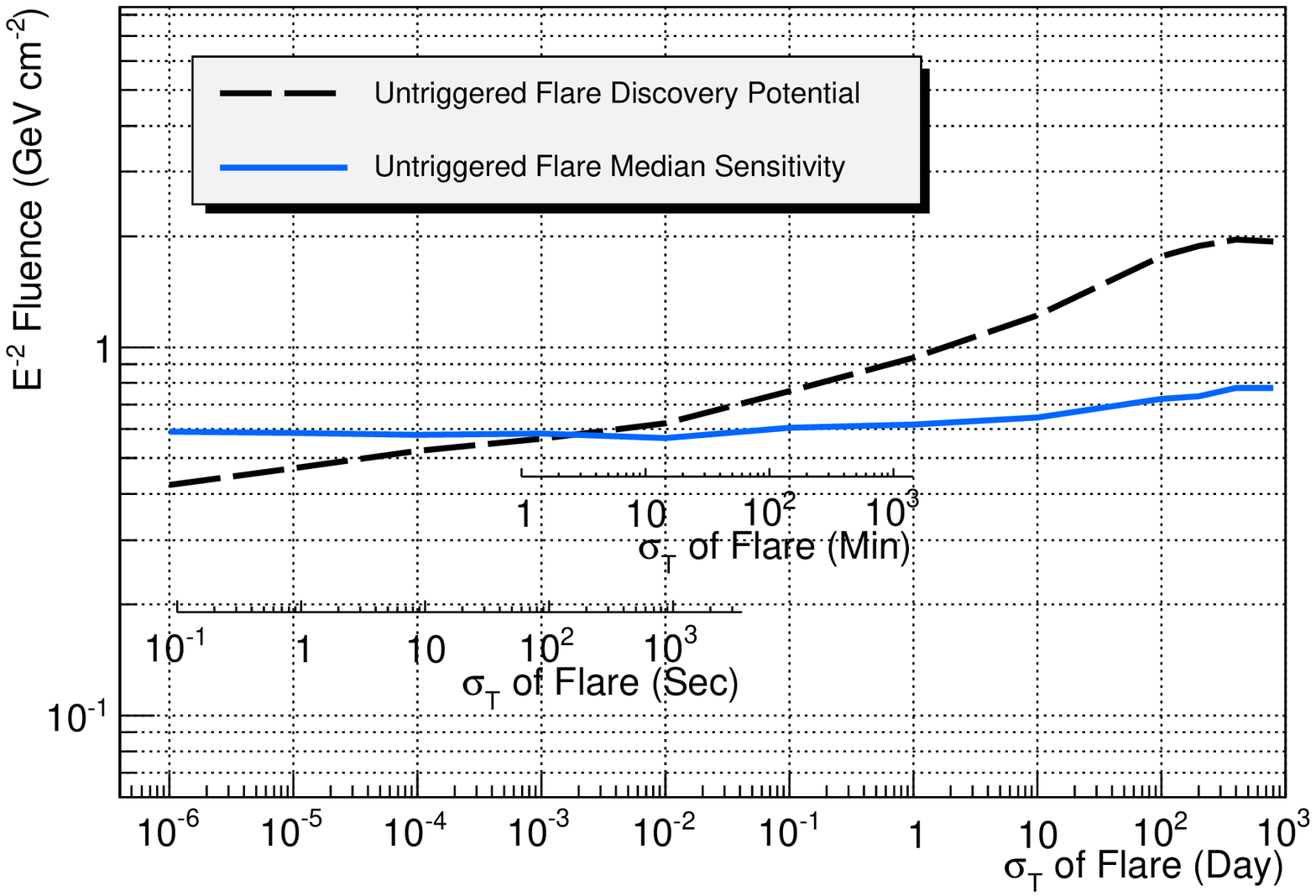}
 \end{tabular}
 }
 \caption{The 50\% 5$\sigma$ discovery potential and 90\% median sensitivity in terms of the mean number of events (left) and fluence (right) for a fixed source at $+16^{o}$ declination. The number of events for the median sensitivity and discovery potential for the time-integrated search are also shown. Flares with a $\sigma_T$ of less than 100 days, or a FWHM of less than roughly half the total livetime, have a better discovery potential than the steady search.}
 \label{fig6.1}
\end{figure}

As described in \citet{Braun:2009wp}, the numerical maximizer needs an initial candidate flare (a ``first guess''). In that paper this first guess was obtained by selecting events within 5$^\circ$ of the source location (in this analysis we use the criteria ${\mathcal{S}_i}/{\mathcal{B}_i}>1$, where $\mathcal{S}_i$ and $\mathcal{B}_i$ are defined in Section \ref{sec4}, omitting the time term). The data is broken up into sets of $m$ temporally consecutive events, where 2 $\le m \le$ 5, for initial testing.
So, for a stream consisting of time-ordered events numbered 1,2,3,4,5,6,7..., the initial test uses events (1,2), (2,3), (3,4) etc., (1,2,3), (2,3,4), (3,4,5) etc., (1,2,3,4), (2,3,4,5), (3,4,5,6) etc., and (1,2,3,4,5), (2,3,4,5,6) etc.
Each set is tested using the described likelihood formula for compatibility with a flare with an E$^{-2}$ spectrum. The candidate with the best test statistic (from Equation \ref{marginalize}) is used as the initial first guess in the maximization. 
In the current analysis, the maximum number of consecutive events for the initial test has been increased with respect to \citet{Braun:2009wp} to m = 10, improving the sensitivity to longer flares. This brings the performance of the analysis close to that of the corresponding time-integrated analysis at large time scales. Given that more than 5 events are required for discovery for $\sigma_T > 2$ days (see Fig~\ref{fig6.1}), if the maximum is not increased the method will occasionally only find a subset of the injected events, hence increasing the total signal required to cross the threshold for discovery.

Figure~\ref{fig6.1} (left) shows the mean number of injected events from a Gaussian time function needed for a $5\sigma$ discovery for 50\%
cases (black solid line) as a function of the duration of the flare $\sigma_T$ for a fixed source location at declination of $+16^{\circ}$, though sources at other declinations yield similar results. This is compared to the 
number of events needed in a time-integrated search (black dashed line): the number of events needed to discover a flare of 1s duration is about a factor of 4 lower than for a time-integrated search. 
At long timescales the flare search performs only 10\% worse than the time-integrated search, even with 2 additional free parameters in the fit.
In the same plot the median upper limits at 90\% c.l. are shown for the time-dependent search and for the time-integrated one.
On the right the corresponding fluence is given, where a correction is introduced for the median dead time during a given flare as a function of the flare width (see Figure~\ref{fig6.0}).

The fact that the 50\% 5$\sigma$ discovery potential curve descends below the 90\% median upper limit curve is due to the effect of Poisson statistics. 
The untriggered search must observe at least two events in order to identify a flare.  For a simulated flaring source which injects a mean number of events $\mu$, $\mu$ must equal at least 1.68 for 50\% of simulated trials corresponding to 2 or more signal events.
Therefore, at the shortest timescales, the mean signal needed for a discovery in 50\% of trials asymptotically approaches 1.68 events.
We find the sensitivity at 90\% CL saturates at 2.9 events, which is already near the time-independent sensitivity of 3.15 events and the statistical limit. This is the reason why the discovery potential curve is lower than the sensitivity in Figure~\ref{fig6.1}.

The method is applied as an all-sky scan over a grid ($0.5^{\circ} \times 0.5^{\circ}$) in right ascension and declination. The final result of the analysis is the set of best fit parameters from the location with the highest test statistic value. A final $p$-value for this analysis is obtained by performing the same scan on scrambled data sets, and counting the fraction of scrambled sets which have a maximum test statistic greater than or equal to the maximum found in the data. 

\subsection{Results}

Using the 40-string data, the location which deviates most from the background expectation is found at (RA,Dec)=(254.75$^{\circ}$, +36.25$^{\circ}$). Two events are found, with a best-fit spectrum $\hat{\gamma}$ of 2.15, mean of the flare $\hat{T_o}$ of MJD 54874.703125 and width $\hat{\sigma_T}$ of 15 seconds. The two events are 2.0$^{\circ}$ apart in space and 22 seconds apart in time. The $-\log_{\mathrm{10}}$($p$-value) corresponding to this observation is 4.67. A clustering of higher significance is seen in 56\% of scrambled skymaps (276 out of 500), a result consistent with the null hypothesis of background-only data.

\begin{figure*}[hp]
 \centering{
   \includegraphics[width=5.0in]{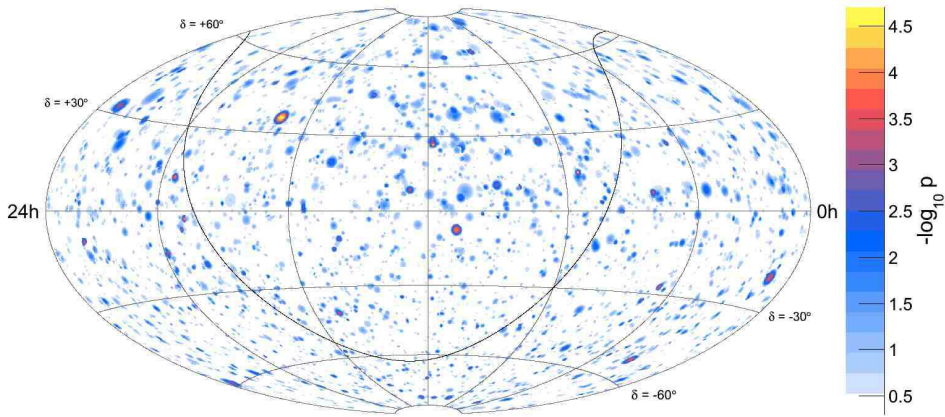}
 }
 \caption{The equatorial coordinate map shows the $p$-value of the most significant flare in time and space at each location of the grid where the likelihood is calculated. The $p$-value is indicated on the z-scale on the right. The black curve is the Galactic plane.
}
\label{fig5p}
\end{figure*}

\begin{figure*}[hp]
 \centering{
   \includegraphics[width=5.0in]{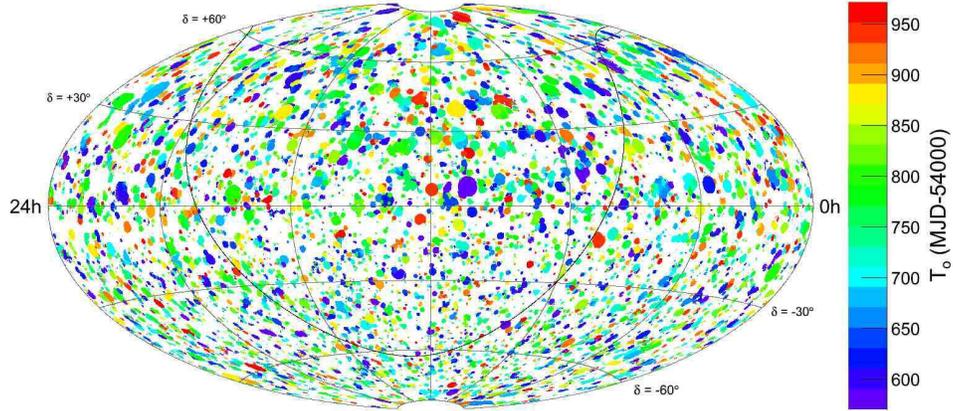}
 }
 \caption{The equatorial coordinate map shows the best fit of the mean time of the flare $\hat{T_o}$ (MJD-54,000) for the most significant flare found at each location of the grid where the likelihood is calculated. The black curve is the Galactic plane.
}
\label{fig5m}
\end{figure*}

\begin{figure*}[h]
 \centering{
   \includegraphics[width=5.0in]{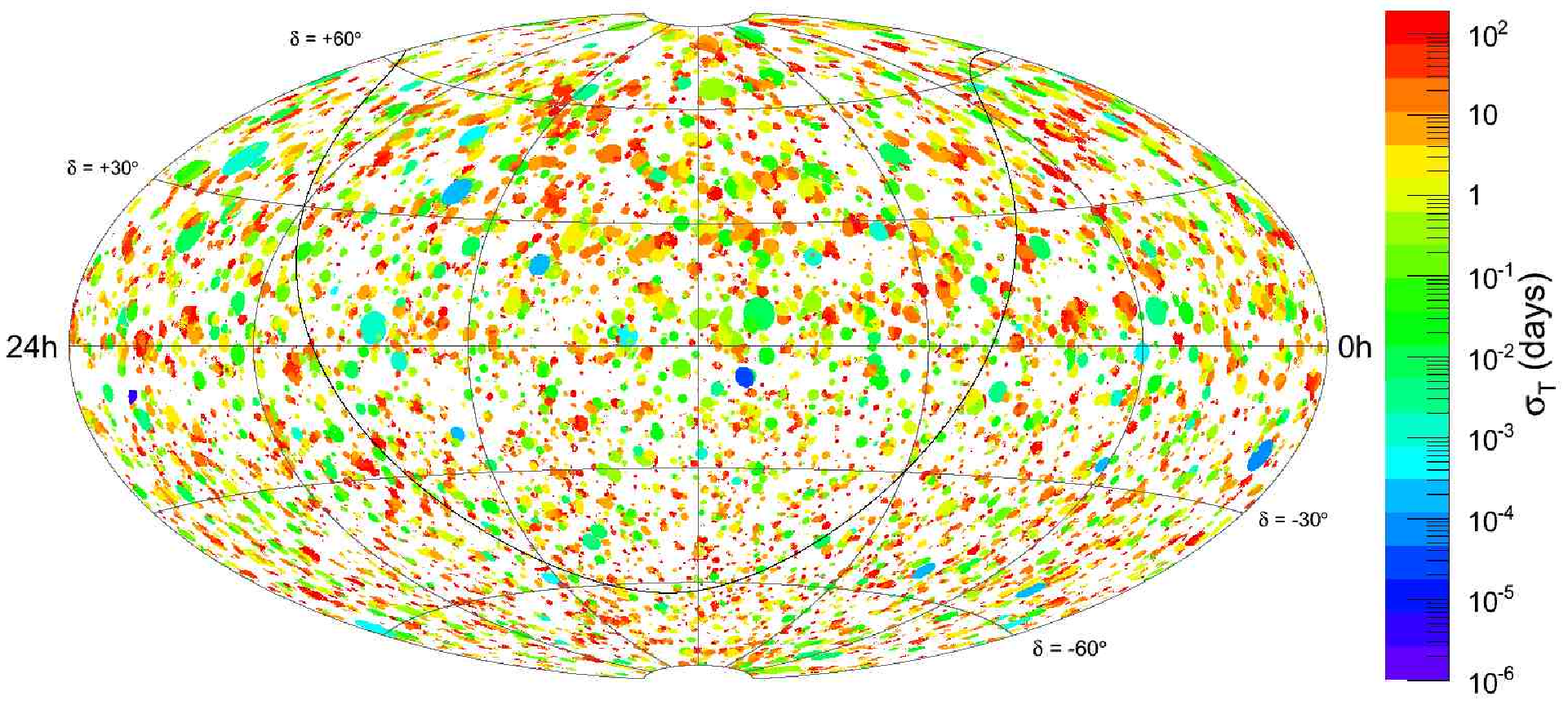}
 }
 \caption{The map in equatorial coordinates of the best fit width $\hat{\sigma_T}$, in days, of most significant flare at a given location found at each location of the grid where the likelihood is calculated in the search. The black curve is the Galactic plane.
}
\label{fig5w}
\end{figure*}

Figures \ref{fig5p} to \ref{fig5w} show maps of the pre-trial $p$-values and best-fit parameters $\hat{T_o}$ and $\hat{\sigma_T}$. Figures \ref{fig5m} and \ref{fig5w} require that the best-fit number of signal events be greater than zero, white area corresponds to being consistent with no flare being detected.
A large effective trial factor of $2.6\times10^4$ is generated by scanning the whole sky. Therefore it is desirable to look only at a few sources in order to decrease the trials, which is done in Section \ref{sec5}.

\section{Time Scan for Candidate Sources}
\label{sec5}

By targeting specific, a priori promising directions in the sky, an analysis can reduce the effective trial factor of the all-sky scan. 
One way this can be done is by performing the analysis in Section \ref{sec6} at the specified locations only. Here, this search was instead implemented using a time-clustering algorithm developed in \citet{Satalecka:2007zz} and \citet{BazoAlba:2009we}, which achieves similar performance. The algorithm finds the most significant flare in a period by testing the most promising time windows, which are defined by the times of the neutrino events.

For the source list, variable bright astrophysical objects from the entire sky are selected. Sources are taken from the Fermi-LAT Bright Source List \citep{Abdo:2009mg}, whose data taking period overlaps with the IceCube 40-string sample. The sources include 30 blazars (24 FSRQs and 6 BL Lacs), one high-mass X-ray binary, one radio galaxy and seven unidentified objects. In this analysis the following selection criteria are defined for choosing the most promising variable astrophysical sources:
\begin{itemize}
\item Classified as variable by Fermi-LAT,
\item Flux [100 MeV - 1 GeV] $>1.1 \times 10^{-7}$ photons cm$^{-2}$s$^{-1}$.
\end{itemize}

The definition of variability provided by the Fermi-LAT Bright Source list is that the 
observation has a probability of less than 1\% of being a steady source
(i.e. variability flag=T). The second requirement sets a minimum photon flux, motivated
by the correlation between neutrinos and photons emitted from the
source predicted by hadronic models. The average photon flux from 100 MeV to 1 GeV of all Fermi variable sources is 2.3 $\times10^{-7}$ photons cm$^{-2}$s$^{-1}$. The flux threshold chosen keeps 60\% of these sources. The list of selected candidates contains 40 objects (see Table~\ref{table_ic40list_results}), 18 in the southern sky and 22 in the northern sky.

\subsection{Method and Expected Performance}

For a given source location, signal-like events are defined as having a time-integrated ${\mathcal{S}_i}/{\mathcal{B}_i}>1$, where $\mathcal{S}_i$ and $\mathcal{B}_i$ are defined in Section \ref{sec4}, omitting the time term.
Each pair of these event times assigns a starting and ending time ($t_i$ and $t_j$, respectively), to the flare search windows $\Delta T_{\mathrm{ij}} = t_j-t_i$. 
The longest flare duration is constrained in the algorithm to be 30 days.
Apart from this constraint the algorithm loops over the events, testing all windows defined by each set of signal-like events specified above, maximizing for the signal fraction and spectrum for each set. The test statistic is calculated for the most promising flare time windows and the best time window (i.e. the highest test statistic value) is chosen as a flare candidate.

The signal time probability $S^{\mathrm{time}}$ is defined by:

\begin{equation}
S^{\mathrm{time}}_{i}(t_i,t_j)=\frac{1}{\Delta T_{ij}} ~~.
\label{bg_time}
\end{equation}
The time probability $S^{\mathrm{time}}_{i}$ is constant, no particular time structure is assumed during the flare. The statistics of signals expected will be small enough that no particular functional form should be discernable.

\begin{figure*}[ht]
 \centering{
   \includegraphics[width=4.5in]{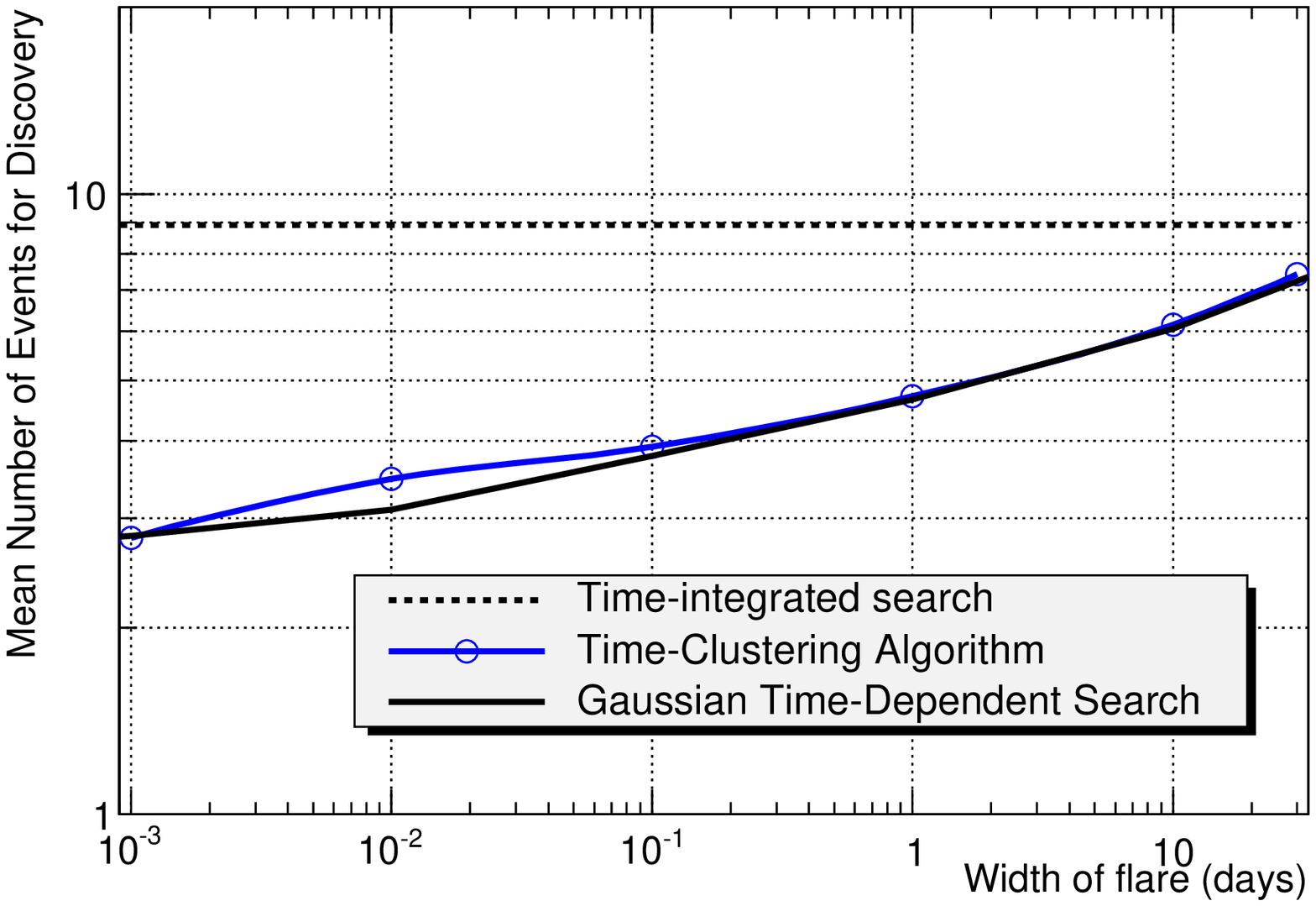}
 }
 \caption{Mean number of events for a 5$\sigma$ discovery in 50\% of trials as a function of the flare duration, calculated as an example for a point source at dec=$16^{\circ}$ ra=$343^{\circ}$ using the time-clustering method (this section) and the method from section \ref{sec4}.}
 \label{plot_ic40_discpot_vs_t}
\end{figure*}

The mean number of events needed for this analysis to achieve a 5$\sigma$ discovery with 50\% of trials was calculated for different widths of simulated flares (see Figure~\ref{plot_ic40_discpot_vs_t}). The flare duration was investigated in the range from 30 days to 20$\mu$s, the minimum time between events. The discovery potential is very similar to the method of Section~\ref{sec4}, which is also plotted in Figure~\ref{plot_ic40_discpot_vs_t}. For flares with duration on the order of minutes, one third of the events needed in a time integrated search is necessary for a detection with the untriggered flare method. 

\begin{figure}[t]
\centering{
\includegraphics[width=4.5in]{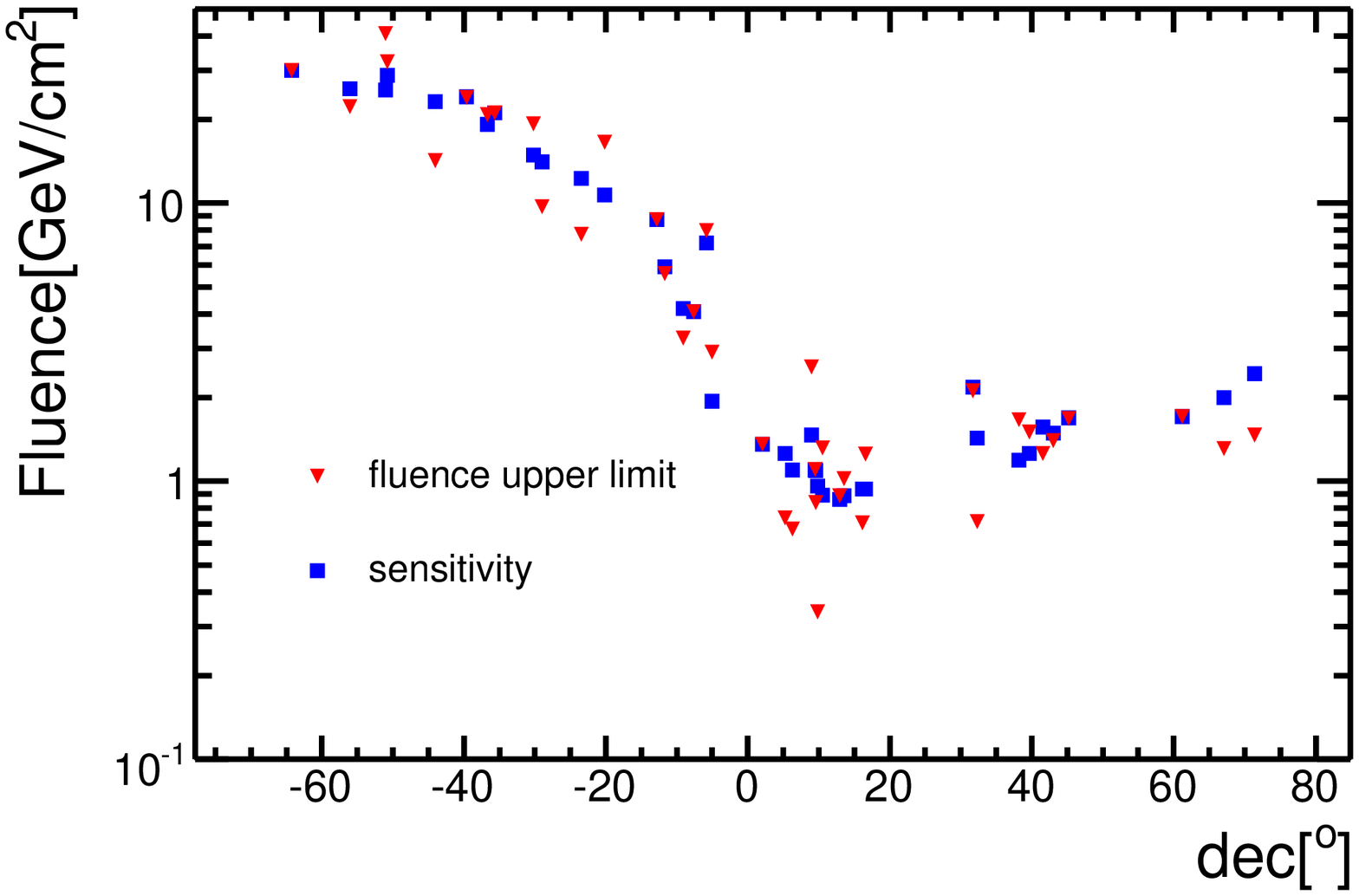}}
\caption[Fluence upper limits of the 40 selected sources as function
of declination.]{Fluence upper limits (red triangles) of the most significant cluster for each of the 40 selected sources calculated in the time windows given in Table \ref{table_ic40list_results} versus declination. The blue squares represent the median sensitivity on the fluence calculated for the same time windows.
}
\label{plot_ic40_unblinding_results}
\end{figure}

\subsection{Results}

The time-scan looking for neutrino flares was applied to the 40 selected
source candidates using IceCube 40-string data. No significant excess
above the atmospheric background is found.
The results and upper limits for each source are
presented in Table~\ref{table_ic40list_results} and summarized
in Figure \ref{plot_ic40_unblinding_results}.
The highest fluctuation observed corresponds to 0FGL J0643.2+0858 (dec=$8.9^{o}$,
ra=$100.8^{o}$) with a $p$-value of 7\% (1.5 $\sigma$) pre-trial. The corresponding best time cluster was 14.3 days, lasting from MJD 54846.5 to 54860.8. Correcting for the trial factor from looking at 40 sources, the final post-trial $p$-value is 95\%. 
The post-trial $p$-value is obtained from the distribution of the maximum test statistic for many equivalent samples obtained by scrambling the time of events.

\begin{landscape}
\begin{center}
\begin{longtable}{@{}|c|c|c|c|c|c|c|c|c|c|@{}}
\caption{Results for pre-defined variable astrophysical source candidates using the likelihood time clustering algorithm.}
\label{table_ic40list_results}\\
\hline
Source & Other Name &  dec [$^\circ$] & ra [$^\circ$]& $\log E_{\mathrm{min}}$ & $\log E_{\mathrm{max}}$ &  $p$-value & $\Delta t$ & $T_\mathrm{start}$ &  Fluence Limit \\
(0FGL)&  &  &  & (GeV) & (GeV) &   & (days) & (MJD)  & (GeV/cm$^{2}$) \\
\hline 
\endfirsthead
\multicolumn{9}{c}%
{{\bfseries \tablename\ \thetable{} -- continued from previous page}} \\
\hline
Source & Other Name &  dec [$^\circ$] & ra [$^\circ$]& $\log E_{\mathrm{min}}$ & $\log E_{\mathrm{max}}$ &  $p$-value & $\Delta t$ & $T_\mathrm{start}$ &  Fluence Limit \\
(0FGL)&  &  &  & (GeV) & (GeV) &   & (days) & (MJD)  & (GeV/cm$^{2}$) \\

\hline
\endhead
\hline
\endfoot
\hline \hline
\endlastfoot

J1123.0-6416 &              & -64.27 & 170.8 & 5.7 & 7.8 & 0.89 & 4.7 & 54718.1 & 30.1 \\
J1328.8-5604 &              & -56.08 & 202.2 & 5.7 & 7.8 & 0.59 & 0.66 & 54641.2 & 22.2 \\
J0210.8-5100 & PKS 0208-512 & -51.01 & 32.71 & 5.7 & 7.8 & 0.11 & 12.9 & 54750.1 & 40.6 \\
J0910.2-5044 &              & -50.74 & 137.6 & 5.7 & 7.8 & 0.38 & 1.53 & 54585.7 & 32.2 \\
J0538.8-4403 & PKS 0537-441 & -44.06 & 84.72 & 5.7 & 7.8 & 0.75 & 3 & 54833.7 & 14.2 \\
J1802.6-3939 &              & -39.66 & 270.7 & 5.7 & 7.8 & 0.97 & 16.4 & 54576.3 & 24.1 \\
J0229.5-3640 & PKS 0227-369 & -36.68 & 37.38 & 5.7 & 7.8 & 0.42 & 4.15 & 54798.4 & 20.8 \\
J1457.6-3538 & PKS 1454-354 & -35.64 & 224.4 & 5.6 & 7.8 & 0.96 & 7.17 & 54788.9 & 21.2 \\
J2158.8-3014 & PKS 2155-304 & -30.24 & 329.7 & 5.6 & 7.8 & 0.15 & 0.124 & 54620.6 & 19.3 \\
J1746.0-2900 &              & -29    & 266.5 & 5.6 & 7.9 & 0.71 & 10.4 & 54934.3 & 9.69 \\
J0457.1-2325 & PKS 0454-234 & -23.43 & 74.29 & 5.5 & 7.9 & 0.72 & 7 & 54890.1 & 7.73 \\
J1911.2-2011 & PKS 1908-201 & -20.19 & 287.8 & 5.4 & 7.9 & 0.1 & 6.45 & 54696.1 & 16.6 \\
J1813.5-1248 &              & -12.8  & 273.4 & 5.2 & 7.9 & 0.5 & 4.34 & 54899.7 & 8.72 \\
J0730.4-1142 & PKS 0727-11  & -11.71 & 112.6 & 5.1 & 7.9 & 0.53 & 0.882 & 54866.6 & 5.59 \\
J1512.7-0905 & BZQ J1512-0905 & -9.093 & 228.2 & 4.8 & 7.8 & 0.65 & 13 & 54855.9 & 3.27 \\
J2025.6-0736 & PKS 2022-07  & -7.611 & 306.4 & 4.6 & 7.7 & 0.91 & 2.08 & 54622.6 & 4.07 \\
J1256.1-0547 & 3C 279       & -5.8   & 194   & 4.2 & 7.6 & 0.45 & 21.8 & 54944.4 & 7.95 \\
J0017.4-0503 &              & -5.054 & 4.358 & 4.0 & 7.6 & 0.16 & 3.25 & 54734.9 & 2.91 \\
J1229.1+0202 & 3C 273       & 2.045  & 187.3 & 3.7 & 7.3 & 0.96 & 28.5 & 54562.7 & 1.36 \\
J1015.9+0515 & PMN J1016+0512 & 5.254 & 154  & 3.7 & 7.0 & 0.74 & 4.46 & 54915.7 & 0.74 \\
J1830.3+0617 &              & 6.287 & 277.6  & 3.7 & 6.9 & 0.71 & 26.2 & 54624.3 & 0.675 \\
J0643.2+0858 &              & 8.983 & 100.8  & 3.7 & 6.7 & 0.07 & 14.3 & 54846.5 & 2.57 \\
J2147.1+0931 & PKS 2144+092 & 9.519 & 326.8  & 3.7 & 6.6 & 0.49 & 27.8 & 54737.9 & 1.1 \\
J1751.5+0935 & OT 081       & 9.591 & 267.9  & 3.7 & 6.6 & 0.66 & 6.22 & 54917.2 & 0.84 \\
J2327.3+0947 & PKS 2325+093 & 9.794 & 351.8  & 3.7 & 6.6 & 0.82 & 2.84 & 54603.2 & 0.339 \\
J1504.4+1030 & PKS 1502+106 & 10.51 & 226.1  & 3.7 & 6.6 & 0.17 & 6.47 & 54777.4 & 1.32 \\
J1553.4+1255 & PKS 1551+130 & 12.92 & 238.4  & 3.7 & 6.5 & 0.48 & 0.488 & 54566.3 & 0.885 \\
J0531.0+1331 & PKS 0528+134 & 13.53 & 82.76  & 3.6 & 6.4 & 0.38 & 0.581 & 54597 & 1.02 \\
J2254.0+1609 & 3C 454.3     & 16.15 & 343.5  & 3.6 & 6.3 & 0.66 & 7.94 & 54594.4 & 0.71 \\
J0238.6+1636 & AO 0235+164  & 16.61 & 39.66  & 3.6 & 6.3 & 0.12 & 0.216 & 54776.6 & 1.25 \\
J1522.2+3143 & TXS 1520+319 & 31.73 & 230.6  & 3.4 & 6.0 & 0.52 & 3.95 & 54869 & 2.11 \\
J1310.6+3220 & B2 1308+32   & 32.34 & 197.7  & 3.4 & 5.9 & 0.76 & 27.7 & 54671.8 & 0.716 \\
J1635.2+3809 & 4C +38.41    & 38.16 & 248.8  & 3.3 & 5.8 & 0.092 & 0.268 & 54795 & 1.66 \\
J1641.4+3939 &              & 39.67 & 250.4  & 3.3 & 5.8 & 0.29 & 0.268 & 54795 & 1.5 \\
J0320.0+4131 & NGC 1275     & 41.52 & 50     & 3.3 & 5.8 & 0.62 & 28.3 & 54576.7 & 1.26 \\
J0222.6+4302 & 3C 66A       & 43.04 & 35.65  & 3.3 & 5.7 & 0.51 & 19.6 & 54641.1 & 1.4 \\
J0654.3+4513 & B3 0650+453  & 45.22 & 103.6  & 3.3 & 5.7 & 0.93 & 5.85 & 54903.5 & 1.69 \\
J0240.3+6113 &              & 61.23 & 40.09  & 3.1 & 5.5 & 0.91 & 6.19 & 54699.5 & 1.71 \\
J1849.4+6706 & S4 1849+67   & 67.1  & 282.4  & 3.1 & 5.3 & 0.7 & 10.6 & 54708.9 & 1.31 \\
J0722.0+7120 & S5 0716+71   & 71.35 & 110.5  & 3.0 & 5.3 & 0.75 & 7.11 & 54864.7 & 1.47 \\

\hline
\end{longtable}
\vspace{0.3cm}
{\small Note.- The source name is the 0FGL catalogue designation. The $p$-value was calculated from simulated background skymaps, $\Delta t$ is the flare duration of the best cluster and $T_\mathrm{start}$ its starting time. The fluence upper limit was calculated by integrating $d\Phi/dE \times E$ over the declination dependent energy range from $E_{\mathrm{min}}$ to $E_{\mathrm{max}}$ to contain 90\% of signal spectrum and $\Delta t$, assuming a neutrino energy spectrum of $E^{-2}$.}
\end{center}
\end{landscape}

\section{Triggered Searches Based on Continuous Photon Observations}
\label{sec7}

When there is specific timing information about the activity of an astronomical object, that information can be used to perform a targeted search with reduced background. This section describes searches in which the photon observations are essentially continuous, and this complete set of flux measurements in time is used. For flares lasting on the order of one day, MWL information can produce a discovery with about one third fewer signal events with respect to untriggered searches (\citet{Braun:2009wp}).

The source selection was motivated by Fermi alerts, which are only issued for sources seen at a flux level greater than $ 2 \times 10^{-6}$ photons s$^{-1}$ cm$^{-2}$. The selected sources are listed in Table \ref{fermiflaresresult}.
These sources include 6 FSRQs, one BL Lac and one unidentified object. The lightcurves were produced for this work using the Fermi Public Release data, using the diffuse class event selection. For each source the \emph{Fermi} Science Tools v9r15p2 package is used to select photons from within $2^{\circ}$ of each source and calculate the total exposure. Photon events with zenith angles greater than 105$^{\circ}$ were excluded to avoid contamination due to the Earth's albedo. Time bins of one day width were then used to calculate an average flux.
There are two modifications to this procedure: the blazar 3C 454.3 was seen in a massive outburst before official science operation \citep{Abdo:2009ef}, for this source the published lightcurve is used. Also, the source PKS 1502+106 was noted to have a large outburst immediately before official science operations began, extending several days after the public information begins \citep{2008ATel.1650....1C,Ciprini:2009kr}.
PKS 1502+106 is taken to be flaring since the time of the alert at a fixed flux level. 
This flaring activity is a possible orphan flare in hard X-rays, since the SWIFT-BAT did not observe any evident flare in the 15-50 keV band while SWIFT XRT and UVOT observed a flare in soft X-rays and optical.

\subsection{Method and Expected Performance}

A Maximum Likelihood Block (MLB) algorithm \citep{Scargle:1997jp,Resconi:2009sw} is used to denoise the lightcurves by iterating over the data points to select periods from the lightcurves which are consistent with constant flux once statistical errors are taken into account.
The MLB algorithm compares the likelihood that a set of points between $x_i$ and $x_f$ is compatible at a confidence level with the change of state between the points $x_c$ and $x_{c+1}$ contained in the interval.
The confidence level requires that for a given set of data points from $x_i$ to $x_f$ that the set of points from $x_i$ and $x_c$ and $x_{c+1}$ to $x_f$ be:

\begin{equation}
\log{\frac{\mathcal{L}(x_i, x_c)\mathcal{L}(x_{c+1}, x_f)}{\mathcal{L}(x_i, x_f)}} > \log{C}~, 
\end{equation}
where $\mathcal{L}(x_a, x_b)$ represents the likelihood that a set of points from $x_a$ to $x_b$ represents a constant flux state from the source between the points, and $C$ is the confidence level. The level of the flux state is determined using the error-weighted mean of the points tested. The method iterates over the different possible changepoints $x_c$, taking the most likely changepoint for the entire dataset and iterating over each subsection of the data. We tested values of $\log{C}$ from 1 to 1000, for values below 5, the typical denoised lightcurve typically follow each data point, for values from 9 to 100 very similar results for the denoised lightcurves were found. The final value of $\log{C}$ used in the analysis was 20.

With the hypothesis that the intensity of the neutrino emission follows the intensity of the photon lightcurve, the signal time PDF is simply the normalized lightcurve itself. A slightly modified hypothesis is that the neutrino emission follows the lightcurve, but only when the photon flux goes above a certain threshold $F_{\mathrm{th}}$. In this case, the value of $F_{\mathrm{th}}$ can be used as a free parameter in the analysis, finding the value of the threshold which maximizes the significance of the data. This method also avoids any penalty from making an incorrect {\emph a priori} choice on a flaring threshold.
$F(t_i)$ is defined as the value of the denoised lightcurve at $t_i$ and the fit parameter $F_{\mathrm{th}}$ is the flux threshold below which no neutrino emission is assumed (i.e. $S^{\mathrm{time}}_{i}$=0 if $F(t_i)\le F_{\mathrm{th}}$). In the case of $F(t_{i})\ge F_{\mathrm{th}}$, the probability of neutrino emission is assumed to be proportional to the flux level above that threshold:

\begin{figure}[ht]
 \centering{
   \includegraphics[width=6.5in]{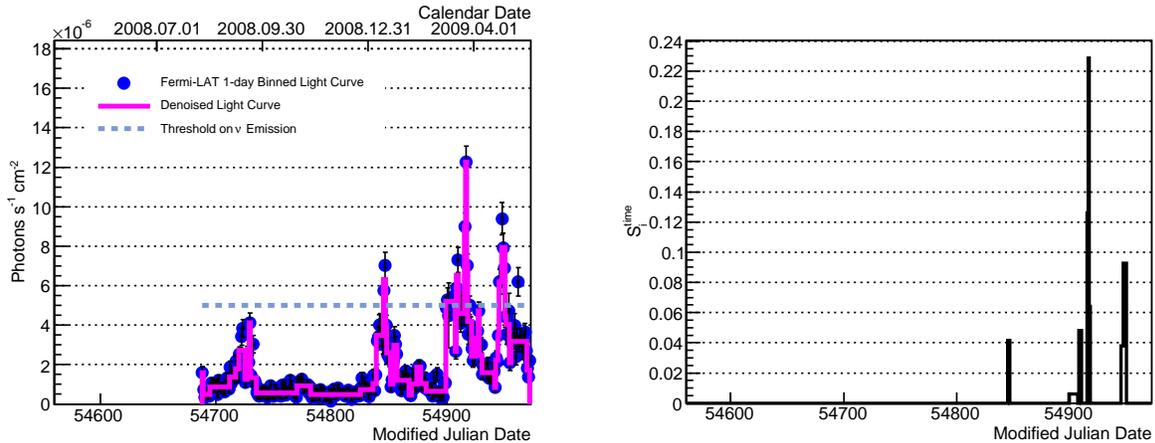}
 }
   \caption{(Left) An example of the one-day binned Fermi lightcurve (blue points, with statistical errors) and denoised lightcurve (pink solid line) for the blazar PKS 1510-089. The dashed line is an example fit threshold. The lightcurve begins here on August 10, 2008 (MJD 54688), when Fermi science operations began, while the time axis shows the entire 40-string data taking period. (Right) The time PDF used in the neutrino signal hypothesis corresponding to the example photon threshold shown in the left graph ($5\times10^{-6}$ photons s$^{-1}$ cm$^{-2}$).}
\label{figpks1510}
\end{figure}

\begin{equation}
S^{\mathrm{time}}_{i} = \frac{F(t_{i})-F_{\mathrm{th}}}{N_{f}};
\end{equation}
where the normalization factor $N_{f}$ is the integral of the denoised lightcurve above the threshold. This time-dependent PDF is then used as before in Equation \ref{llh_signal_time1}. This method is illustrated in Figure~\ref{figpks1510}.

\begin{figure*}[h]
 \centering{
   \includegraphics[width=6.5in]{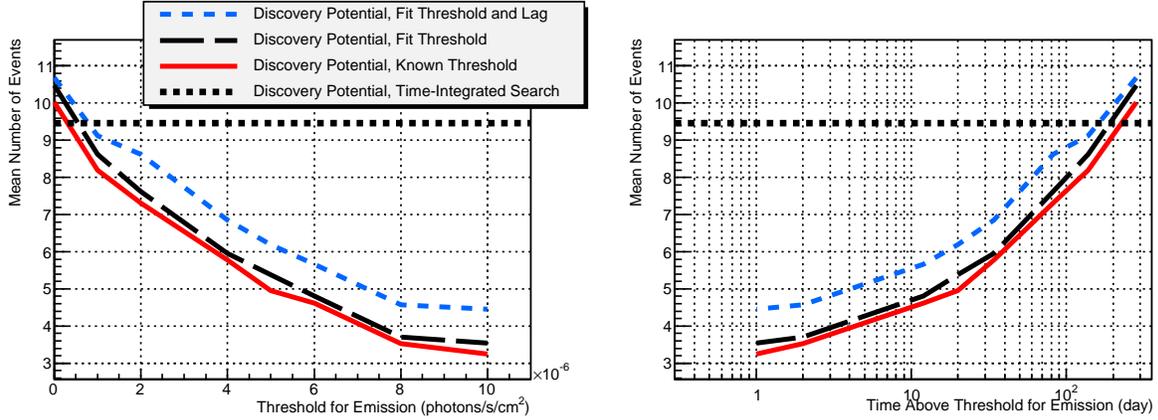}
 }
 \caption{The plot of the $5\sigma$ 50\% discovery potential for the source PKS 1510-089 (the corresponding lightcurve is shown in Figure~\ref{figpks1510}), as a function of the true flux threshold for neutrino emission (left) and as a function of the duration the lightcurve spends above the threshold (right). The discovery potential curves are plotted for the time-integrated case (black short dashed line), and from bottom to top for the case where the threshold is fixed to the true threshold (solid red line), 
the case where the threshold is a free parameter (black long dashed line, used in this analysis) and the case where there is an unknown lag (up to $\pm$ 50 days) between GeV and neutrino emission (blue dashed line).}
\label{discocomp.pks1510}
\end{figure*}

The effect of adding this additional degree of freedom by fitting for $F_{\mathrm{th}}$ can be seen in Figure~\ref{discocomp.pks1510}. The effect is small compared to the penalty of fixing the threshold to an incorrect value.
The effect of allowing an unknown lag up to $\pm50$ days between the photon and neutrino emissions was also tested, and was found to give a marked increase in the number of events required for discovery. Hence, we used the method allowing only up to a $\pm 0.5$ day lag that accounts for the 1 day binning of lightcurves.

\subsection{Results}
\label{sec7.2}

The results from all sources are listed in Table \ref{fermiflaresresult}. The most significant source is PKS 1502+106, which has a pre-trial $p$-value of 5\%. 
 The method finds one high-energy event during the August 2008 flare. The prescription to provide the post-trial $p$-value was to consider the most significant among the flares obtained in this and Section \ref{sec8} with the 40-string configuration.
 The post-trial $p$-value is 29\%, which is compatible with background fluctuations.

\begin{sidewaystable}[!h]
  \begin{center}
    \begin{tabular}{ |c|c|c|c|c|c|c|c|c|}
    \tableline
    Source & dec [$^\circ$] & ra [$^\circ$]& $\log E_{\mathrm{min}}$ & $\log E_{\mathrm{max}}$ & pre-trial & Best-Fit Threshold                  &  Duration  & Upper limit\\
           &  &  & (GeV) & (GeV) & $p$-value & (10$^{-6}$ cm$^{-2}$ s$^{-1}$) &  (days)  & (GeV/cm$^{2}$) \\
    \hline
    PKS 1510-089    & $-9.1$  & 228.2 & 4.8 & 7.8 & ---   & 0     & 282  & 2.49  \\
    3C 66A/B        & 43.0    & 35.7  & 3.3 & 5.7 & 0.47  & 0.675 & 57   & 0.778 \\
    3C 454.3        & 16.2    & 343.5 & 3.6 & 6.3 & 0.20  & 9.47  & 2.5  & 0.422 \\
    PKS 1454-354    & $-35.6$ & 224.4 & 5.6 & 7.8 & ---   & 0     & 282  & 13.1  \\
    3C 279          & $-5.8$  & 194.0 & 4.2 & 7.6 & 0.47  & 2.34  & 6    & 1.69  \\ 
    PKS 0454-234    & $-23.4$ & 74.3  & 5.5 & 7.9 & ---   & 0     & 282  & 7.87  \\
    PKS 1502+106    & 10.5    & 226.1 & 3.7 & 6.6 & 0.049 & 3.13  & 8    & 0.370 \\   
    J123939+044409  & 4.7     & 189.9 & 3.7 & 7.1 & ---   & 0     & 282  & 0.661 \\
    \tableline
    \end{tabular}
  \end{center}
  \caption{Sources tested with the 40 string data and pre-trial $p$-values for the flare search with continuous lightcurves. In the event of an under-fluctuation no $p$-value is calculated and there is no indication of flaring behavior. The overlap between the Fermi public release data and the 40-string data taking period is 282 days. The duration column corresponds to the amount of time for the lightcurve which is above the best-fit threshold.}
  \label{fermiflaresresult}
\end{sidewaystable}

\section{Triggered Searches Based on Intermittent Photon Observations}
\label{sec8}

Ground based observatories such as H.E.S.S., MAGIC, and VERITAS cannot monitor sources continuously, because they can only operate when there is good nighttime visibility.  Their observations are nevertheless extremely important for neutrino searches, because they detect photons at TeV energies that are potentially better correlated to neutrinos of the energies to which IceCube is sensitive.
While these observatories can issue alerts for source activity, they often cannot constrain the beginning or end of the activity to within a few days.
For alerts such as these, the present analysis uses a simple time cut, taking a window for events one day before and after the identified flare. The catalogue corresponding to the 40-string data includes a recent suspect ``orphan flare'' at the level of 10 Crab from Mrk 421 observed by VERITAS and MAGIC \citep{collaboration:2009gga,Vittorini:2008nf}. During the 40-string period, there was no overlap between sources with flares seen by Fermi and by IACTs and tested in this paper.

\subsection{Method and Expected Performance}

The nature of this analysis is a simple cut in time between $t_{\mathrm{min}}$ and $t_{\mathrm{max}}$, which can be expressed as:

\begin{equation}
S^{\mathrm{time}}_i = \frac{H(t_{\mathrm{max}}- t_i) \times H(t_i -t_{\mathrm{min}})}{t_{\mathrm{max}}-t_{\mathrm{min}}}
\end{equation}

where $t_i$ is the arrival time of the event, $t_{\mathrm{max}}$ and $t_{\mathrm{min}}$ are the predefined the upper and lower bounds of the time window defining the flare, and $H$ is the Heavyside step function. This time-dependent signal PDF is then used in Equation \ref{llh_signal_time1}. In this analysis, the signal population size $n_s$ and spectrum index $\gamma$ are the only fit parameters.

\subsection{Results}

Five of the seven flares tested with the 22 string data (Table \ref{ic22flare}) showed no excess of events in the vicinity of the corresponding sources in the selected time windows, while S5 0716+71 and 1ES 1959+650 showed one event each. The post-trial $p$-value is 14\%, the most significant flare being the 10 day flare of S5 0716+71. This result is consistent with background fluctuations.

\begin{sidewaystable}[!h] 
 \begin{center}
  \begin{tabular}[!t]{|c|c|c|c|c|c|c|c|c| }
   \hline
   Source & dec [$^\circ$] & ra [$^\circ$] &  Alert Ref. &	Time & $\log E_{\mathrm{min}}$ & $\log E_{\mathrm{max}}$ & $p$-value      & Fluence\\
          & & &             &   Window       & (GeV) & (GeV) & (pre-trial)  & Limit  \\
         & & &             &   (MJD)       &         &       &              & \footnotesize{(GeV/cm$^{2}$)} \\
   \hline
    1ES 1959+650 & 65.1 & 300.1 & \footnotesize{\citep{Bottacini}}   & 54428-54433         & 3.1 & 5.4 &  ---   & 1.78 \\
    1ES 1959+650 & 65.1 & 300.1 & \footnotesize{\citep{Whipple1es1959}} & 54435.5-54440.5  & 3.1 & 5.4 &  0.08  & 1.81 \\
    3C 454.3 & 16.2 & 343.5 & \footnotesize{\citep{Vercellone:2008uf}} & 54305-54311       & 3.8 & 6.5 &  ---   & 0.812 \\
    3C 454.3 & 16.2 & 343.5 & \footnotesize{\citep{Vercellone:2008ht}} & 54416-54426       & 3.8 & 6.5 &  ---   & 0.812 \\
    Cyg X-1 & 35.2 & 299.5 & \footnotesize{\citep{GCN6745}}          & 54319.5-54320.5     & 3.4 & 5.0 &  ---   & 1.19 \\
    S5 0716+71 & 71.3 & 110.5 & \footnotesize{\citep{Chen:2008fu}}   & 54350-54356         & 3.1 & 5.3 &  ---   & 2.00 \\
    S5 0716+71 & 71.3 & 110.5 & \footnotesize{\citep{Chen:2008fu}}   & 54392-54402         & 3.1 & 5.3 &  0.02  & 2.00 \\
  \hline
  \end{tabular}
  \end{center}
  \caption{Flare list during the 22 strings data-taking: source name, references for the alert, interval in modified Julian day, pre-trial $p$-value. The $p$-value is calculated only when $\hat{n}_s$ is greater than zero.}
 \label{ic22flare}
\end{sidewaystable}

Of the six sources tested with the 40-string data (Table \ref{ic40flare}), five showed no excess of events in the vicinity of the sources during the selected time periods. 
The final post-trial $p$-value for the 40 string analysis (considering the 15 flares in Table \ref{fermiflaresresult} and Table \ref{ic40flare}) is 29\%.

\begin{sidewaystable}[!h] 
\begin{center}
\begin{tabular}{|c|c|c|c|c|c|c|c|c| }
 \hline
  Source & dec [$^\circ$] & ra [$^\circ$] &  Alert Ref. &	Time & $\log E_{\mathrm{min}}$ & $\log E_{\mathrm{max}}$ & $p$-value      & Fluence\\
          & & &             &   Window       & (GeV) & (GeV) & (pre-trial)  & Limit  \\
         & & &             &   (MJD)       &         &       &              & (GeV/cm$^{2}$) \\
   \hline
    Markarian 421 & 38.2 & 166.1  & \citep{Pichel:2009}      & 54586-54592 & 3.3 & 5.8 & ---   & 1.51 \\
                  &  &  & \citep{Vittorini:2008nf}           & 54621-54631 &      &      &      &      \\
    W Comae 	  & 28.2 & 185.4 & \citep{Acciari:2009xz}    & 54623-54627 & 3.4 & 6.0 & ---   & 1.32 \\
    S5 0716+714   & 71.3 & 110.5 & \citep{Mazin:2009}        & 54572-54582 & 3.0 & 5.3 & 0.34  & 3.26 \\
    SGR 0501+4516 & 45.3 & 75.3 & \citep{2008ATel.1683....1P}         & 54700-54706 & 3.2 & 5.7 & ---   & 1.72 \\
    1ES 1218+304  & 30.2 & 185.3 & \citep{Acciari:2010xu}    & 54859-54864 & 3.4 & 6.0 & ---   & 1.43 \\
    Markarian 501 & 39.8 & 253.5 & \citep{Pichel:2009}       & 54951-54953 & 3.3 & 5.8 & ---   & 1.78 \\
  \hline
  \end{tabular}
  \end{center}
  \caption{Flare list seen with occasional coverage during the 40-string data-taking. References are for the alert which prompted the selection. The $p$-value is calculated only when $\hat{n}_s$ is greater than zero. The flare windows for Markarian 421 were added together, only one $p$-value and upper limit for both periods is calculated.}
  \label{ic40flare}
\end{sidewaystable}

\section{Conclusions}
\label{sec10}

In this paper we discuss four time-dependent searches: two are ``untriggered'' 
and scan over direction, energy and time to find clusters of neutrino events; 
the others are ``triggered'' by multi-wavelength information. While the first two 
searches are generic and sensitive to flares not seen in photons, the others are 
more sensitive because of the reduced trial factor but concern specific catalogues 
of variable sources. Time-dependent searches can be more sensitive to short 
flares thanks to the reduction of the background of atmospheric muons and neutrinos 
over short time scales. The untriggered search using a predefined catalogue of 40 
variable sources shows a post-trial $p$-value of 95\% and upper limits on the 
fluence are calculated (see Table \ref{table_ic40list_results}).
Eight of these sources are also triggered on observed flares in the triggered search.
The all-sky scan over all directions finds that the most significant cluster of events is separated 
in time by 22 s and in space by $2^{\circ}$ and has a $p$-value of 56\%. 
The most significant observation of flaring from 14 sources in catalogues compiled using Fermi-LAT 
and IACT alerts during the 40-string configuration data taking is PKS 1502+106, 
with a $p$-value of 29\% after trials. The most significant flare triggered by MWL observations during the 
22-string configuration is S5 0716+71 with a $p$-value of 14\% after trials. All 
these results are compatible with a fluctuation of the background.

The complete IceCube detector began taking data in April 2011, and is expected to have an effective area twice that of 
40 strings at high energies, and up to a factor of ten at 100 GeV. This is especially
a boon to time-dependent neutrino point source analyses, which are limited by the
statistics of the signal events observed. Time-dependent analyses using data from the full detector will have
roughly a factor of two to three improvement in the fluence upper limits and discovery 
potentials from point sources, a significant improvement in the capabilities of
neutrino astronomy.

\acknowledgments

We acknowledge the support from the following agencies:
U.S. National Science Foundation-Office of Polar Programs,
U.S. National Science Foundation-Physics Division,
University of Wisconsin Alumni Research Foundation,
the Grid Laboratory Of Wisconsin (GLOW) grid infrastructure at the University of Wisconsin - Madison, the Open Science Grid (OSG) grid infrastructure;
U.S. Department of Energy, and National Energy Research Scientific Computing Center,
the Louisiana Optical Network Initiative (LONI) grid computing resources;
National Science and Engineering Research Council of Canada;
Swedish Research Council,
Swedish Polar Research Secretariat,
Swedish National Infrastructure for Computing (SNIC),
and Knut and Alice Wallenberg Foundation, Sweden;
German Ministry for Education and Research (BMBF),
Deutsche Forschungsgemeinschaft (DFG),
Research Department of Plasmas with Complex Interactions (Bochum), Germany;
Fund for Scientific Research (FNRS-FWO),
FWO Odysseus programme,
Flanders Institute to encourage scientific and technological research in industry (IWT),
Belgian Federal Science Policy Office (Belspo);
University of Oxford, United Kingdom;
Marsden Fund, New Zealand;
Japan Society for Promotion of Science (JSPS);
the Swiss National Science Foundation (SNSF), Switzerland;
A.~Gro{\ss} acknowledges support by the EU Marie Curie OIF Program;
J.~P.~Rodrigues acknowledges support by the Capes Foundation, Ministry of Education of Brazil.

\bibliographystyle{apj}



\end{document}